\documentclass[11pt,a4paper]{article}
\usepackage[utf8]{inputenc}
\usepackage[bottom=1.0in,top=1.0in]{geometry}
\usepackage{amsmath}
\usepackage{enumitem}
\usepackage{amsfonts}
\usepackage{mathtools}
\usepackage{amssymb}
\usepackage{breqn}
\clearpage
\usepackage{verbatim} %This package used for multiline comment
\usepackage[utf8]{inputenc}
\usepackage[english]{babel}
\usepackage{multicol}
\begin{document}
%\begin{comment}

\begin{center}
\textbf{
Analytic evaluation of some three- and four- electron atomic integrals involving s STO's and exponential correlation with unlinked $r_{ij}$'s
}
\end{center}

\begin{center}
\textbf{B Padhy}\\
\end{center}

\begin{center}
Former faculty member, Department of Physics, Khallikote College, \\Brahmapur - 760 001, Odisha, India.\\E-mail : bholanath.padhy@gmail.com
\end{center}

\begin{abstract}
The method of evaluation outlined in a previous work has been utilized here to evaluate certain other three- electron and four- electron atomic integrals involving s Slater-type orbitals and exponential correlation with unlinked $r_{ij}$'s. Limiting expressions for various such integrals have been derived, which has not been done earlier. Closed-form expressions for $<r_{12} r_{13} / r_{14}>$, $<r_{12}r_{34}/r_{23}>$, $<r_{12}r_{23}/r_{34}>$, $<r_{12}r_{13}/r_{34}>$ and $<r_{12}r_{34}/r_{13}>$ have been obtained.
\end{abstract}

\begin{flushleft}
\end{flushleft}
\hfill \break
\textbf{1. Introduction}\\
\hfill \break
It is well accepted that the details of the way in which electrons mutually correlate their motion in a many-electron system are to be included in order to obtain very accurate wave functions and energies for the system by employing quantum mechanical calculations in the framework of the Rayleigh - Ritz variational procedure [1].  The two standard methods which include these electron - electron correlations are (i) the configuration - interaction (CI) method and (ii) the Hylleraas (Hy) method. In the CI method, the trial function is represented as a linear  combination of a large number of antisymmetrized products of one- electron functions, each product referring to one particular configuration of the system. This method is easy to apply to any system, in principle, for calculations. However, it is plagued with the weakness of extremely slow convergence. On the other hand, the Hy method, in which the interelectronic separation coordinates are included explicitly in variational basis functions, as was first proposed by Hylleraas [2, 3], gives quick convergence in energy compared to the CI method. This method is regarded as the most powerful method among the existing theoretical approaches to produce results of high accuracy [4]. It has been quite successful in obtaining highly accurate energy for two-electron systems [5, 6]. However, the number of interelectronic separation coordinates involved in the trial function for an N-electron atom is N(N-1)/2, and hence  the evaluation of the corresponding generating integrals in the Hy method becomes more and more difficult for systems with increasing N. As far as the knowledge of the author goes, the application of the Hy method for variational calculations is limited upto only four - electron systems till date [7]. 

As regards the application of the Hy method to three-electron atomic systems, James and Coolidge [8] were the first to attempt to compute the energy for the ground state of Li atom. They were successful in expressing the so called triangle integral (with integrand involving all the three interelectronic separation coordinates) in terms of auxiliary functions A, V and W, which are themselves of one-, two-, and three-dimensional integrals, respectively. For knowledge about progressive development of various numerical methods of evaluation, with greater accuracy, of three-electron correlated integrals over atomic Slater-type orbitals (STO's) [9], the reader is advised to go through the review article by King [10], the paper by Pelzl and King [11] and references therein, as well as the paper by Yan and Drake [12]. However, analytic expressions have been reported by Frolov and Smith [13] for the functions A, V and W. In all reports cited above in this paragraph, no exponential correlation has been considered.

In an alternative approach, employing Fourier transform method, Fromm and Hill [14] could succeed for the first time in obtaining a closed-form expression for the triangle integral involving exponential correlation. The expression reported by them [14] does not involve the auxiliary  functions A, V and W. Subsequently, five more reports of analytic evaluation of the triangle integral with certain comments and modifications were published by different authors [15 - 19]. Employing the Hylleraas basis set with and/or without exponential correlation, many investigations relating to various properties of the three-electron atomic system have been reported [20-23] by Pachucki and Puchalski and coworkers. Making use of results in [14, 24], formulas for the recursive generation of many other three-electron exponentially correlated  integrals have been reported by Harris [25].

For the accurate determination of wave functions and energies for atomic systems with more than three electrons by avoiding computational difficulty met in the Hy method, an alternative procedure was systematically developed in early seventies by Sims and Hagstrom [26, 27] by introducing explicitly interelectronic separation coordinates into a CI wave function with certain restrictions.  In this method, known as the Hylleraas - configuration - interaction (Hy-CI) method, each term in the expansion of the proposed CI function was restricted to contain explicitly only one two-electron correlation factor of the form $r_{ij}^p$, with $p$ restricted to the value 0, 1 and 2. Of course, with $p=0$, one gets the original CI wave function. To include electron-electron correlations in this manner in the wave function expansion had been first proposed by James and Coolidge [8], and was later employed by others [28-31]. The Hy-CI method was employed for the first time by Sims and Hagstrom [26, 27] for the study of the ground state of the beryllium atom by taking a 107 - configuration wave function. An auxiliary function X, which is itself a four-dimensional integral, was introduced by them in addition to the auxiliary functions A, V and W introduced in [8]. Also, a computational scheme was reported by them for the accurate calculation of the auxiliary function X in terms of A, V and W functions. The Hy-CI method was further utilized successfully to investigate the ground state and some excited states of Li atom [32-34], the ground state of $Li^-$ ion [35], the ground state of neon atom [36], and the ground state of neutral helium and He-like ions [37]. Several low-lying states in Li atom and $Be^+$ ion have been investigated by Ruiz et al. [38] recently employing the Hy-CI analysis. The most accurate result available using this approach for the ground state of Be atom and its isoelectronic sequence yields energies accurate to better than one microhartree[39,40]. Sims and Hagstrom, in a series of papers [41-43], have discussed thoroughly certain mathematical and computational science issues in high precision calculation of the three-electron triangle integral, three-electron kinetic energy and four-electron integrals which arise while employing Hy-CI method of variation. Some three-electron and four-electron integrals have been evaluated by Ruiz [44,45] by integration over the coordinates of one electron and calculated to a high degree of accuracy. Same method of evaluation was employed by Ruiz[46,47] to calculate the two-electron kinetic energy and the three-electron kinetic energy. Frolov et al. [48] have investigated bound state spectra and properties of the doublet states in $Li$ atom and some $Li$-like ions employing Hy-CI and CI methods. The basic four-electron atomic correlated integral with integrand involving all the six eletron-electron separation coordinates was reduced to a sum of several auxiliary functions X (denoted as $W_4$) [49-51], and later reevaluated by King [52] to increase the effectiveness in computation and reported closed-form expressions for some integrals as special cases. Analytical expressions for the auxialiary functions A, V, W and X (redenoted as $A_1$,  $A_2$, $A_3$ and $A_4$, respectively) along with their highly accurate values have been reported in [53, 54]. A computationally efficient and numerically stable method was reported in [4,55] for the highly accurate calculation of auxiliary functions W and X (denoted as $W_3$ and $W_4$). All the papers cited above in this paragraph do not involve exponential correlation. Further, it is observed that the trial wave function expansion has been chosen to consist of a large number of terms even in the Hy and the Hy-CI variation methods. For example, in the latest investigation for the ground state of $Be$ atom by Hy method a 200 term Hylleraas  wave function was taken in [7]; for the investigation of the ground state of $Be$ atom and its isoelectronic sequence by Hy-CI method, about 40000 terms were considered in [39,40].

It is expected that with the involvement of exponential correlation in the Hy and the Hy-CI methods, the convergence will be quicker even with less number of terms, though the evaluation of respective integrals will be relatively difficult. Accordingly, one speaks of the Extended-Hylleraas-configuration-interaction (E-Hy-CI) method [56] in which each configuration in the CI wave function expansion is restricted to contain at best one correlation factor of the form $r_{ij}^{\nu} exp (-\lambda_{ij}r_{ij})$. Thus closed-form expressions have been reported by the author [57] for certain three- and four- electron atomic integrals which involve exponential correlation and s STO's with $r_{ij}$'s having unlinked indices; also, three different five- electron atomic integrals of this category have been evaluated in closed-form and reported in [58]. Analytical expressions for some such three- and four- electron atomic integrals have been recently reported by King [59], wherein stability issues for obtaining correct numerical values from analytic formulas were discussed. Certain such three- and four- electron atomic integrals, but involving nonspherically symmetric STO's also evaluated and reported by Wang at al. [56]. There are several earlier reports relating to evaluation of correlated atomic integrals involving exponential correlation [60-64] and without exponential correlation [30,65-68].

The plan of this paper is as follows. In section 2, the key integral to be utilized several times in the paper has been evaluated in closed-form. Certain two- electron atomic integrals have been evaluated in section 3. Three equivalent three- electron atomic integrals represented by two-vortex diagrams have been evaluated in closed-form in section 4. In section 5.1, analytic expressions for four equivalent four- electron atomic integrals represented by three- vortex diagrams, as well as for $<r_{12} r_{13} / r_{14}>$ have been derived. Closed-form expressions for eight different four- electron atomic integrals represented by open square diagrams as well as for $<r_{12} r_{34} / r_{23}>$, $<r_{12} r_{23} / r_{34}>$ , $<r_{12} r_{13} / r_{34}>$ and $<r_{12} r_{34} / r_{13}>$, have been obtained in section 5.2. Limiting expressions have also been derived for various integrals. Concluding remarks are given in section 6.

\hfill\break
\textbf{2. Evaluation of the key integral}\\
\hfill\break
In the previous paper [57], hereinafter referred as paper I, a simple method was outlined for obtaining closed-form expressions for some two-, three-, and four- electron atomic integrals involving spherically symmetric s STO’s and exponential correlation, with the restriction that the inter-electron separation coordinates with unlinked indices only are present in the integrand. This successful analytic evaluation is consistent with the conjecture 'A' put forward by Bonham [63]. The key to this successful evaluation lies in deriving a closed - form expression for the following integral J defined by 
%\end{comment}
%\begin{multicols*}{2}
{
\begin{equation} % Equation - 1
{
J(\lambda_t, \lambda_{st}, r_s) = \int{d\overrightarrow{r_t}(r_tr_{st})^{-1} exp(-\lambda_tr_t - \lambda_{st}r_{st}), 
}}
\end{equation}
where $\overrightarrow{r_s}$ and $\overrightarrow{r_t}$, respectively, are the position vectors of the $s^{th}$ and the $t^{th}$ electrons with respect to the nucleus assumed to be infinitely heavy and situated at the origin of the coordinate system chosen. Here $r_{st}=r_{ts}=|\overrightarrow{r_s}-\overrightarrow{r_t}|$ is the distance between the $s^{th}$ and the $t^{th}$ electrons, and $\lambda_t$ and $\lambda_{st}$ are the exponential parameters. Obviously, $\lambda_{st}=\lambda_{ts}$.

To evaluate the integral in  equation(1), the following Fourier representation 

\begin{equation} % Equation - 2
\frac{exp(-\lambda_{st}r_{st})}{r_{st}} = \frac{1}{2\pi^2}\int {d\overrightarrow{K}\frac{exp[i\overrightarrow{K}\cdot(\overrightarrow{r_s} - \overrightarrow{r_t})]}{K^2 + \lambda_{st}^2}}
\end{equation}
is employed on the right hand side, then the orders of integration are interchanged and the integration over $\overrightarrow{r_t}$ is performed. Subsequently, the integral over the Fourier transform variable $\overrightarrow{K}$ is evaluated by making use of the inverse Fourier transform to get 

\begin{equation}% Equation - 3
J(\lambda_t, \lambda_{st}, r_s) =\frac{4\pi}{\lambda_t^2 - \lambda_{st}^2} \frac{exp(-\lambda_{st} r_s) - exp(-\lambda_tr_s)}{r_s}\cdot
\end{equation}
If $\lambda_{st} \rightarrow \lambda_t$, then L'Hospital's rule for $0/0$ can be employed on the right hand side expression in  equation(3) to obtain

\begin{equation} % Equation - 4 
J(\lambda_t, \lambda_t, r_s) = \frac{2\pi}{\lambda_t}  exp(-\lambda_t r_s)\cdot
\end{equation}
\hfill \break
\textbf{3. Evaluation of two-electron atomic integrals }\\
\hfill \break
\textbf{3.1 The general integral} \\
\hfill \break
The general two-electron atomic integral involving s STO's and exponential correlation, denoted by $I_2$, is given by 

\begin{equation} %Equation - 5
I_2 (i,j,k; \lambda_1, \lambda_2, \lambda_{12}) = \int {d\overrightarrow{r_1} d\overrightarrow{r_2} r_1^i r_2^j r_{12}^k exp(-\lambda_1r_1 - \lambda_2r_2 - \lambda_{12}r_{12})},
\end{equation}
where the integers $i, j, k$ are each $\geqslant -1$, and the exponential parameters $\lambda_1, \lambda_2$ and $\lambda_{12}$ are such that $\lambda_1 + \lambda_2$, $\lambda_1 + \lambda_{12}$ and $\lambda_2 + \lambda_{12}$ are positive, although there are no such restrictions on these parameters individually [62]. The graphical representation [63] of the integral in  equation(5) is just a straight line connecting the electron positions 1 and 2, indicating that the integrand involves only one inter-electron coordinate $r_{12}$ which takes into account correlation.  equation(5) can be written as \\

%Equation - 6
\begin{equation}
\begin{split}
&I_2(i, j, k; \lambda_1, \lambda_2, \lambda_{12}) = \left(-\frac{\partial}{\partial\lambda_1}\right)^{i+1} \left(-\frac{\partial}{\partial\lambda_2}\right)^{j+1}\left(-\frac{\partial}{\partial\lambda_{12}}\right)^{k+1} \\
\kern 4pc \times &I_2 (-1, -1, -1 ; \lambda_1, \lambda_2, \lambda_{12}),
\end{split}
\end{equation}
where $I_2(-1, -1, -1 ; \lambda_1, \lambda_2, \lambda_{12})$, in which $i=j=k=-1$, is termed as the corresponding generating integral and is denoted, in short, by $I_2^g (\lambda_1, \lambda_2, \lambda_{12})$, the superscript 'g' signifying the generating integral. Clearly we have
\begin{equation}%Equation - 7
I_2^g(\lambda_1, \lambda_2, \lambda_{12})=\int {d\overrightarrow{r_1} d\overrightarrow{r_2}(r_1 r_2 r_{12})^{-1} exp(-\lambda_1r_1-\lambda_2r_2-\lambda_{12}r_{12})\cdot
}
\end{equation}
Making use of  equations(1) and (3) in  equation(7) above, it can be recast as 
\begin{equation} % Equation - 8
I_2^g(\lambda_1, \lambda_2, \lambda_{12}) = \int {d\overrightarrow{r_1} r_1^{-1} exp(-\lambda_1r_1) J(\lambda_2,\lambda_{12},r_1),
}
\end{equation} 
and evaluated analytically to obtain the following closed form expression :

\begin{dmath} % Equation - 9
I_2^g(\lambda_1, \lambda_2, \lambda_{12})=16\pi^2 [(\lambda_1+\lambda_2)(\lambda_1+\lambda_{12})(\lambda_2+\lambda_{12})]^{-1} \cdot
\end{dmath} 

The above expression is exactly in agreement with the one obtained by Calais and Lowdin [61] by employing perimetric cordinates, first introduced by Coolidge and James [69]. The result in  equation(9) can be used in  equation(6) to obtain closed - form expressions for a sequence of integrals given by  equation(5) by the method of parametric differentiation. The expressions obtained in this manner for certain three such integrals are given here only for the purpose of record :

\begin{equation}% Equation - 10
\begin{split}
&I_2(-1, -1, 0; \lambda_1, \lambda_2, \lambda_{12})\\&= 16\pi^2(\lambda_1+\lambda_2+2\lambda_{12})      
[(\lambda_1+\lambda_2)(\lambda_1+\lambda_{12})^2(\lambda_2+\lambda_{12})^2]^{-1},
\end{split}
\end{equation} 
\begin{equation}% Equation - 11
\begin{split}
&I_2(0, 0, -1; \lambda_1, \lambda_2, \lambda_{12}) \\&= 32\pi^2[(\lambda_1+\lambda_2+\lambda_{12})^2+\lambda_1\lambda_2][(\lambda_1+\lambda_2)^3(\lambda_1+\lambda_{12})^2(\lambda_2+\lambda_{12})^2]^{-1},
\end{split}
\end{equation} 
\begin{equation}% Equation - 12
\begin{split}
&I_2(0, 0, 0; \lambda_1, \lambda_2, \lambda_{12}) \\&= 64\pi^2[\lambda_1\lambda_2\lambda_{12} + (\lambda_1+\lambda_2+\lambda_{12})^3] [(\lambda_1+\lambda_2)(\lambda_1+\lambda_{12})(\lambda_2+\lambda_{12})]^{-3}\cdot
\end{split}
\end{equation} 
It is observed that the right hand side expressions in  equations(10-12) are symmetric with interchange of $\lambda_1$ and $\lambda_2$, as expected. \\
\hfill \break
\textbf{3.2 Some other two- electron integrals}\\
\hfill \break
It is worth mentioning here four other nonsingular two-electron integrals of interest which do not come under the general category of integrals defined in  equation(5), but are useful for the evaluation of certain relativistic corrections. These are : (i) $I_2(-1, -1, -2 ; \lambda_1, \lambda_2, \lambda_{12})$,(ii)$I_2(-1, -2, -1 ; \lambda_1, \lambda_2, \lambda_{12})$, (iii) $I_2(-2, -1, -1 ; \lambda_1, \lambda_2, \lambda_{12})$ and (iv)$I_2(-2, -2, -1 ; \lambda_1, \lambda_2, \lambda_{12})$. \\
\hfill \break
\textbf{Evaluation of integrals in (i) - (iii)} \\
\hfill \break
Integrals in (i) -(iii) are evaluated by the method of integration with respect to parameters $\lambda_{12}, \lambda_2$ and $\lambda_1$, respectively, and using the expression in  equation(9). For example, to evaluate the integral in (i), one makes use of the following observation :
\begin{equation} % Equation 13
\frac{\partial}{\partial\lambda_{12}}I_2(-1,-1,-2;\lambda_1, \lambda_2, \lambda_{12}) = -I_2^g(\lambda_1, \lambda_2, \lambda_{12}) \cdot
\end{equation}
Inserting  equation(9) in  equation(13) and then integrating both sides with respect to the parameter $\lambda_{12}$ in the range $\lambda_{12}$ to $\infty$, and noting that $I_2(-1,-1,-2;\lambda_1, \lambda_2, \lambda_{12})\rightarrow 0$ as $\lambda_{12}\rightarrow \infty$, the following closed - form expression is obtained :
\begin{align} % Equation - 14
I_2(-1,-1, -2; \lambda_1, \lambda_2, \lambda_{12}) 
= \frac{16\pi^2}{\lambda_2^2 -\lambda_1^2} ln \left(\frac{\lambda_2 +\lambda_{12}}{\lambda_1 +\lambda_{12}}\right),
\end{align} 
which is in conformity with  equation(9) of the report of Puchalski and Pachucki [22]. If $\lambda_{12}\rightarrow 0 $ on both sides of the above equation, it simplifies to 
\begin{align}%Equation - 15
I_2(-1,-1, -2; \lambda_1, \lambda_2, 0)
 = 16\pi^2(\lambda_2^2 -\lambda_1^2)^{-1} ln\left(\lambda_2/\lambda_1\right),
\end{align}
which is identical with  equation(15) of the paper reported by Roberts [65], but evaluated in a different approach. It is worth mentioning here that the integral in  equation(15) above has also been evaluated by us independently by using Hylleraas coordinates [3] to get the right hand side expression exactly, as reported in paper I.

If the parameter $\lambda_2\rightarrow\lambda_1$ in the integral in (i), then the following closed-form expression is obtained, as a special case, by employing L' Hospital's rule for $0/0$ on the right hand side expression in  equation(14) :
\begin{equation} %Equation 16
I_2(-1, -1, -2;\lambda_1, \lambda_1, \lambda_{12}) = 8\pi^2[\lambda_1(\lambda_1+\lambda_{12})]^{-1} \cdot
\end{equation} %Equation 17
If, further, $\lambda_1=\lambda_2=\lambda_{12}=\delta$, then
\begin{equation}
I_2(-1, -1, -2;\delta, \delta, \delta) = \left(2\pi/\delta\right)^2\cdot
\end{equation}

The integral in (ii) is evaluated in the same manner as in (i) to obtain the following closed-form expression :
\begin{equation} %Equation 18
I_2(-1, -2, -1; \lambda_1, \lambda_2, \lambda_{12}) = \frac{16\pi^2}{\lambda_1^2 - \lambda_{12} ^2}ln\left(\frac{\lambda_1+\lambda_2}{\lambda_2+\lambda_{12}}\right)\cdot
\end{equation}
In yet another method, the same integral can be evaluated by making use of  equation(3) and the standard integral [70] 
\begin{equation}%Equation 19
\int_0^\infty \frac{dx}{x}[exp(-ax) - exp(-bx)] = ln\left(b/a\right)
\end{equation}
to establish  equation(18), as was done in paper I. If $\lambda_{12}\rightarrow\lambda_1$, then employing L'Hospital's rule for $0/0$ on the right hand side expression in  equation(18), one gets
\begin{equation} %Equation 20
I_2(-1, -2, -1; \lambda_1, \lambda_2, \lambda_1) = 8\pi^2[\lambda_1(\lambda_1+\lambda_2)]^{-1} \cdot
\end{equation} %Equation 21
If, further, $\lambda_1=\lambda_2=\lambda_{12}=\delta$, then one obtains
\begin{equation}
I_2(-1, -2, -1; \delta, \delta, \delta) = \left(2\pi/\delta\right)^2 \cdot
\end{equation}
Closed-form expression in  equations(20) and (21) can also be obtained directly by carrying out integrations employing  equations(1) and (4).

Following exactly the same two different methods of evaluation employed for the integral in (ii), it is easy to obtain the following closed-form expression for the integral in (iii) : 
\begin{equation} %Equation 22
I_2(-2, -1, -1; \lambda_1, \lambda_2, \lambda_{12}) = \frac{16\pi^2}{\lambda_2^2-\lambda_{12}^2}ln \left(\frac{\lambda_1+\lambda_2}{\lambda_1+\lambda_{12}}\right),
\end{equation}
which is in conformity with  equation(12) of the paper of Harris et al.[62]. As pointed out earlier,  equation(22) can also be obtained from  equation(18) by interchanging $\lambda_1\leftrightarrows\lambda_2$. Similarly, the following equations are established :
\begin{equation} %Equation 23
I_2(-2, -1, -1; \lambda_1, \lambda_2, \lambda_2) = 8\pi^2[\lambda_2(\lambda_1+\lambda_2)]^{-1},
\end{equation}
\begin{equation} %Equation 24
I_2(-2, -1, -1; \delta, \delta, \delta) = \left(2\pi/\delta\right)^2 \cdot
\end{equation}
\hfill \break
\textbf{Evaluation of integral in (iv)}\\
\hfill \break
The expression for the integral in (iv) can be obtained by substituting $\alpha=\lambda_1, \beta=\lambda_2$ and $\gamma=\lambda_{12}$ in the general expression reported in [62]
\begin{equation}%25
I_2(-2, -2, -1; \alpha, \beta, \gamma) = \frac{8\pi^2}{\gamma}Q(\alpha,\beta,\gamma) ,
\end{equation}
where
\begin{multline}%26
Q(\alpha,\beta,\gamma)=\frac{1}{2}ln^2\left(\frac{\alpha+\gamma}{\beta+\gamma}\right)+dilog\left(\frac{\alpha+\beta}{\beta+\gamma}\right)+dilog\left(\frac{\alpha+\beta}{\alpha+\gamma}\right)+\frac{\pi^2}{6}\cdot
\end{multline}
Here the dilogarithm function, denoted as $dilog(x)$, is defined by
\begin{equation}
dilog(x) = \int_1^x\frac{lnt}{1-t}dt=\sum_{n=1}^\infty\frac{(1-x)^n}{n^2},
\end{equation}
with the properties that the series is convergent for $|x-1| \leq 1$ and
\begin{equation}
\frac{d}{dx}dilog (x) = \frac{lnx}{1-x}\cdot
\end{equation}

It is observed that the right hand side of  equation(25) is symmetric with respect to interchange $(\alpha\leftrightarrows\beta)$, as expected. It had been shown [62] that $Q(\alpha,\beta, \gamma) \to 0$ as $\gamma\to 0$, so that the right hand side expression in  equation(25) assumes $0/0$ form. Hence, in the limit  $\gamma \to 0$, L'Hospital's rule for $0/0$ was applied to obtain, as a special case, \\
\begin{equation} %eqn-29
I_2(-2, -2, -1; \alpha, \beta, 0) = \frac{16\pi^2}{\alpha}ln\left(\frac{\alpha+\beta}{\beta}\right)+\frac{16\pi^2}{\beta}ln\left(\frac{\alpha+\beta}{\alpha}\right) \cdot
\end{equation}
The above integral has also been evaluated independently in a different approach to establish  equation(29) as pointed out in paper I. Further, if $\alpha=\beta=\gamma=\delta$, then $Q(\delta, \delta, \delta) = \pi^2/6$, since $ln(1)=dilog(1)=0$, and hence
\begin{equation} %Equation 30
I_2(-2, -2,-1; \delta ,\delta ,\delta) =(4/3)(\pi^4/\delta) \cdot
\end{equation} 
\hfill \break
\textbf{4. Evaluation of three-electron atomic integrals }\\
\hfill\break
\textbf{4.1  Integrals with unlinked indices }\\
\hfill\break
There are three equivalent general three - electron integrals involving s STO's and exponential
correlation, which are graphically represented by three equivalent two-vortex diagrams indicating the presence of only two of the three inter-electron separations of the type $r_{st}$ and $r_{su}, s\neq t \neq u = 1,2,3$, each line emanating from a common electron  site 's' in one such integral. These are denoted as (i)$I_{31}$, (ii)$I_{32}$ and (iii) $I_{33}$.\\
\hfill \break
\textbf{Definition and evaluation of $I_{31},  I_{32}$ and $I_{33}$ } \\
\hfill\break
\textbf{(i)}The first integral denoted by $I_{31}$ is defined as \\
\begin{equation} %equation 31
\begin{split}
&I_{31}(i,j,k;l,m ; \lambda_1,\lambda_2,\lambda_3;\lambda_{12},\lambda_{13})
=\int d\overrightarrow{r_1}d\overrightarrow{r_2}d\overrightarrow{r_3} r_1^i r_2^j r_3^k r_{12}^l r_{13}^m\\
&\kern 6pc  \times exp (-\lambda_1r_1-\lambda_2r_2-\lambda_3r_3-\lambda_{12}r_{12}-\lambda_{13}r_{13})\cdot
\end{split}
\end{equation}
Here i,j,k,l and m are integers, each $\geq -1$ and the values of the exponential parameters $\lambda_1, \lambda_2, \lambda_3, \lambda_{12}$ and $\lambda_{13}$ should be such that the integral converges. The above integral can be written as 
\begin{equation} %equation 32
\begin{split}
&I_{31}(i,j,k;l,m ;\lambda_1,\lambda_2,\lambda_3;\lambda_{12},\lambda_{13})\\
&=\left(-\frac{\partial}{\partial\lambda_1}\right)^{i+1}\left(-\frac{\partial}{\partial\lambda_2}\right)^{j+1}\left(-\frac{\partial}{\partial\lambda_3}\right)^{k+1}\left(-\frac{\partial}{\partial\lambda_{12}}\right)^{l+1}\left(-\frac{\partial}{\partial\lambda_{13}}\right)^{m+1} \\
&\kern 9pc \times I_{31}(-1, -1, -1; -1, -1; \lambda_1,\lambda_2,\lambda_3;\lambda_{12},\lambda_{13}) ,
\end{split}
\end{equation}
where the corresponding generating integral, in which $i=j=k=l=m=-1$, is 
\begin{equation} %equation 33
\begin{split}
&I_{31}(-1,-1, -1;-1, -1;\lambda_1, \lambda_2, \lambda_3;\lambda_{12}, \lambda_{13}) 
= \int {d\overrightarrow{r_1}d\overrightarrow{r_2}d\overrightarrow{r_3}(r_1r_2r_3r_{12}r_{13})^{-1}} \\& \kern 11pc \times exp (-\lambda_1r_1-\lambda_2r_2-\lambda_3r_3-\lambda_{12}r_{12}-\lambda_{13}r_{13})\cdot
\end{split}
\end{equation}
It is denoted, in short, by $I_{31}^g(\lambda_1,\lambda_2,\lambda_3;\lambda_{12},\lambda_{13})$  and expressed as 
\begin{equation}
I_{31}^g=\int d\overrightarrow{r_1}r_1^{-1} exp(-\lambda_1r_1) J(\lambda_2,\lambda_{12},r_1) J(\lambda_3,r_{13},r_1),
\end{equation}
where the J's are the integrals defined in  equation(1) with their closed - form expressions given by  equation(3). Substituting the respective closed - form expressions for the J's in  equation(34), the angular integration over the variable $\overrightarrow{r_1}$ is performed easily employing spherical polar coordinates. Then the standard integral given in  equation(19) is employed to obtain the following closed-form expression for the generating integral $I_{31}^g$ : 
\begin{equation} %35
I_{31}^g=\frac{64\pi^3}{(\lambda_2^2-\lambda_{12}^2)(\lambda_3^2-\lambda_{13}^2)}ln\left[\frac{(\lambda_1+\lambda_3+\lambda_{12})(\lambda_1+\lambda_2+\lambda_{13})}{(\lambda_1+\lambda_{12}+\lambda_{13})(\lambda_1+\lambda_2+\lambda_3)}\right] \cdot
\end{equation}
\hfill \break
\textbf{Limiting expressions for $I_{31}^g$} 

If $\lambda_{12}=\lambda_{13}=0$, then
\begin{equation} %quation 36
I_{31}^g(\lambda_1, \lambda_2, \lambda_3;0,0)=\frac{64\pi^3}{\lambda_2^2\lambda_3^2}ln\left[\frac{(\lambda_1+\lambda_3)(\lambda_1+\lambda_2)}{\lambda_1(\lambda_1+\lambda_2+\lambda_3)}\right] \cdot
\end{equation}
If, further, the exponential parameters $\lambda_1=\lambda_2=\lambda_3=\delta$, then
\begin{equation} %quation 37
I_{31}^g(\delta, \delta, \delta;0,0)=64\pi^3ln(4/3)\delta^{-4} \cdot
\end{equation}
Also it is observed in  equation(35) that if either $\lambda_{12}\rightarrow\lambda_2$ or $\lambda_{13}\rightarrow\lambda_3$ or $\lambda_{12}\rightarrow\lambda_2$ and $\lambda_{13}\rightarrow\lambda_3$ simultaneously, the right hand side expression assumes $0/0$ form, and hence limiting expression can be obtained by applying L'Hospital's rule for $0/0$. Thus, if $\lambda_{12}\rightarrow\lambda_2$ alone, one gets

\begin{equation} %quation 38
I_{31}^g(\lambda_1, \lambda_2, \lambda_3;\lambda_2, \lambda_{13})=\frac{32\pi^3}{\lambda_2(\lambda_3^2-\lambda_{13}^2)}\left[\frac{1}{\lambda_1+\lambda_2+\lambda_{13}}-\frac{1}{\lambda_1+\lambda_2+\lambda_3}\right] \cdot
\end{equation} 

Similarly, if $\lambda_{13}\rightarrow\lambda_3$ alone, one obtains

\begin{equation}  %quation 39
I_{31}^g(\lambda_1, \lambda_2, \lambda_3;\lambda_{12}, \lambda_3)=\frac{32\pi^3}{\lambda_3(\lambda_2^2-\lambda_{12}^2)}\left[\frac{1}{\lambda_1+\lambda_{12}+\lambda_3}-\frac{1}{\lambda_1+\lambda_2+\lambda_3}\right] \cdot
\end{equation}

In case $\lambda_{12}\rightarrow\lambda_1$ and $\lambda_{13}\rightarrow\lambda_3$ simultaneously, then the result is

\begin{equation} %quation 40
I_{31}^g(\lambda_1, \lambda_2, \lambda_3;\lambda_2, \lambda_3)=16\pi^3[\lambda_2\lambda_3(\lambda_1+\lambda_2+\lambda_3)^2]^{-1} \cdot
\end{equation}
Also, if each one of all the five exponential parameters $(\lambda$'s$)$ is equal to $\delta$, then
\begin{equation} %quation 41
I_{31}^g(\delta,\delta,\delta ; \delta,\delta)= (16/9)\pi^3\delta^{-4}\cdot
\end{equation}

It is worth mentioning here that  equations(36-41) can also be directly established starting from  equation(33) and replacing the J's in the integrand in  equation(34) as per  equation(3) and / or  equation(4) as desired in the limits $\lambda_{12}\rightarrow\lambda_2$ and / or $\lambda_{13}\rightarrow\lambda_3$, and then carrying out the integration. This statement has actually been verified by obtaining the expressions in  equations(36-41).

Closed-form expressions for a sequence of integrals given by  equation(31) can be obtained for various values of i,j,k,l and m by parametric differentiation method employing  equations(32) and (35). Thus, all the six entries in the fifth column of table 1 of the paper of Harris et al.[68] are reproduced by differentiating first both sides of  equation(35) with respect to suitable parameters and then setting $\lambda_{12}=\lambda_{13} = 0$ in the final expression. As an example, to reproduce the sixth entry, the following integral is considered : 
\begin{equation} %equation 42
\begin{split}
&(64\pi^3)^{-1}I_{31}(-1,-1,-1; 1,1; \lambda_1, \lambda_2, \lambda_3; \lambda_{12}, \lambda_{13})\\
&\kern 3pc = \frac{1}{64\pi^3}\frac{\partial^2}{\partial\lambda_{12}^2}\frac{\partial^2}{\partial\lambda_{13}^2}I_{31}^g(\lambda_1, \lambda_2, \lambda_3; \lambda_{12}, \lambda_{13}) \cdot
\end{split}
\end{equation}
The above differentiations are performed on the right hand side expression in  equation(35) and then in the final expression $\lambda_{12}=\lambda_{13}=0$ substituted. The resulting expression comes out to be exactly identical with the sixth entry.

Also the expression on the right hand side of  equation(36) is in conformity with the fourth entry. Proceeding in a similar manner, the other four entries are exactly reproduced. 

(ii) The second three-electron integral denoted by $I_{32}$ is defined as \\
\begin{equation} %equation 43
\begin{split}
&I_{32}(i,j,k;l,m;\lambda_1,\lambda_2,\lambda_3;\lambda_{21},\lambda_{23})
=\int {d\overrightarrow{r_1}d\overrightarrow{r_2}d\overrightarrow{r_3}r_1^ir_2^jr_3^kr_{21}^lr_{23}^m }\\
&\kern 7pc \times exp(-\lambda_1r_1-\lambda_2r_2-\lambda_3r_3-\lambda_{21}r_{21}-\lambda_{23}r_{23}) \cdot
\end{split}
\end{equation}
The corresponding generating integral, analogous with definitions in  equations(31-33), is given by 
\begin{equation} %equation 44
\begin{split}
&I_{32}^g(\lambda_1,\lambda_2,\lambda_3;\lambda_{21},\lambda_{23})
=\int {d\overrightarrow{r_1}d\overrightarrow{r_2}d\overrightarrow{r_3}(r_1r_2r_3r_{21}r_{23})^{-1}}\\
&\kern 3pc \times exp(-\lambda_1r_1-\lambda_2r_2-\lambda_3r_3-\lambda_{21}r_{21}-\lambda_{23}r_{23}) \cdot
\end{split}
\end{equation}
It is observed that if a change $(1\leftrightarrows2)$ is performed in  equation(33), and noting that $r_{12}=r_{21}$ and $\lambda_{12}=\lambda_{21}$,  equation(44) is obtained. Accordingly, the following closed-form expression for the integral in  equation(44) is obtained from  the right hand side expression in equation(35) by inspection : 

\begin{equation} %Eqn45
I_{32}^g = \frac{64\pi^3}{(\lambda_1^2-\lambda_{12}^2)(\lambda_3^2-\lambda_{23}^2)} ln\left[\frac{(\lambda_2+\lambda_3+\lambda_{12})(\lambda_1+\lambda_2+\lambda_{23})}{(\lambda_2+\lambda_{12}+\lambda_{23})(\lambda_1+\lambda_2+\lambda_3)}\right] \cdot
\end{equation} 

This expression was also reported earlier in paper I. The integral given in  equation(5) and its closed-form expression given in  equation(9) in the paper of Bonham[63] are exactly reproduced by suitable parametric differentiations of both sides of  equation(45). Also various limiting expressions of the generating integral $I_{32}^g$ above can be obtained following the procedure adopted for deriving  equations(36-41) relating to $I_{31}^g$.

(iii) The third three-electron integral denoted by $I_{33}$ is defined as 
\begin{multline} %equation 46
I_{33}(i,j,k;l,m;\lambda_1,\lambda_2,\lambda_3;\lambda_{13},\lambda_{23})
=\int {d\overrightarrow{r_1}d\overrightarrow{r_2}d\overrightarrow{r_3}r_1^ir_2^jr_3^kr_{13}^lr_{23}^m }\\ 
\times exp(-\lambda_1r_1-\lambda_2r_2-\lambda_3r_3-\lambda_{13}r_{13}-\lambda_{23}r_{23}) \cdot
\end{multline}
The corresponding generating integral is given by 
\begin{equation} %equation 47
\begin{split}
&I_{33}^g(\lambda_1,\lambda_2,\lambda_3;\lambda_{13},\lambda_{23})
=\int {d\overrightarrow{r_1}d\overrightarrow{r_2}d\overrightarrow{r_3}(r_1r_2r_3r_{13}r_{23})^{-1}}\\
&\kern 3pc \times exp(-\lambda_1r_1-\lambda_2r_2-\lambda_3r_3-\lambda_{13}r_{13}-\lambda_{23}r_{23})\cdot
\end{split}
\end{equation}
It is observed that, if we make either a change $(2\leftrightarrows3)$ in  equation(44) or a change $(3\leftrightarrows1)$ in  equation(33), we get  equation(47). Hence, by inspection, the following closed-form expression for $I_{33}^g$ is obtained either from  equation(45) by performing a change $(2\leftrightarrows3)$ or from  equation(35) by making a change $(3\leftrightarrows1)$ :
\begin{equation} %Eqn48
I_{33}^g = \frac{64\pi^3}{(\lambda_2^2-\lambda_{23}^2)(\lambda_1^2-\lambda_{13}^2)} ln\left[\frac{(\lambda_1+\lambda_3+\lambda_{23})(\lambda_3+\lambda_2+\lambda_{13})}{(\lambda_3+\lambda_{23}+\lambda_{13})(\lambda_1+\lambda_2+\lambda_3)}\right] \cdot
\end{equation}
Various limiting expressions for the above generating integral can be obtained as outlined in the case of other two generating integrals. \\
\hfill\break
\textbf{4.2 Integral with linked indices }\\ 
\hfill\break
The only other three-electron integral which is graphically represented by a triangle, hence known as the triangle integral, indicates the involvements of all the three inter-electron separation coordinates with linked indices in the integrand. It does not come under the category of integrals mentioned in section 2.1 above.

For the development of various numerical and analytical methods of evaluation, with greater accuracy, of the triangle integral, it is advisable to go through the second and the third paragraph in the introduction.\\
\hfill\break
\textbf{5. Evaluation of four-electron integrals with unlinked indices } \\
\hfill \break
Four-electron integrals with exponential correlation involving only three inter-electron separation coordinates with unlinked indices are divided into two different categories on the basis of their graphical representation ; the first category corresponds to three-vortex diagrams and the second to open squares.\\
\hfill \break
\textbf{5.1  Integrals represented by three-vortex diagrams } \\
\hfill \break
There are four equivalent general four-electron integrals with exponential correlation, each one of which is graphically represented by a three-vortex diagram indicating that the integrand involves explicitly only three inter-electron separation coordinates (of the type $r_{st}, r_{su}$ and $r_{sv}, s\neq t\neq u \neq v =1,2,3,4 $, each line emanating from a common electron site 's') with unlinked indices out of the total six inter-electron separation coordinates of a four-electron atom. These are denoted here as  (i) $I_{41}$, (ii) $I_{42}$,(iii) $I_{43}$ and (iv) $I_{44}$, and defined by
\begin{equation} %Eqn49
\begin{split}
&I_{41}(i, j, k, l; m, n, p; \lambda_1, \lambda_2, \lambda_3, \lambda_4 ; \lambda_{12}, \lambda_{13}, \lambda_{14}) \\
&= \int {d\overrightarrow{r_1}d\overrightarrow{r_2}d\overrightarrow{r_3}d\overrightarrow{r_4}r_1^ir_2^jr_3^kr_4^lr_{12}^m r_{13}^n r_{14}^p} \\
&\times exp(-\lambda_1r_1-\lambda_2r_2-\lambda_3r_3-\lambda_4r_4-\lambda_{12}r_{12}-\lambda_{13}r_{13}-\lambda_{14}r_{14}),
\end{split}
\end{equation}
\begin{align} %Eqn50
\begin{split}
&I_{42}(i, j, k, l; m, n, p; \lambda_1, \lambda_2, \lambda_3, \lambda_4 ; \lambda_{21}, \lambda_{23},\lambda_{24})\\
&= \int {d\overrightarrow{r_1}d\overrightarrow{r_2}d\overrightarrow{r_3}d\overrightarrow{r_4}r_1^ir_2^jr_3^kr_4^lr_{21}^m r_{23}^n r_{24}^p} \\
&\times exp(-\lambda_1r_1-\lambda_2r_2-\lambda_3r_3-\lambda_4r_4-\lambda_{21}r_{21}-\lambda_{23}r_{23}-\lambda_{24}r_{24}),
\end{split}
\end{align}
with similar definitions for the integrals $I_{43}$ and $I_{44}$. Here i,j,k,l,m,n and p are the integers, each $\geq-1$, and the values of the exponential parameters $(\lambda's)$ should be such that the integral converges. Each one of these integrals can be expressed in terms of the corresponding generating integrals in which $(i=j=k=l=m=n=p=-1)$ by the method of parametric differentiation. For example, 
\begin{flalign} %Equation 51
\begin{split}
&I_{41}(i, j, k, l; m, n, p; \lambda_1, \lambda_2, \lambda_3, \lambda_4 ; \lambda_{12}, \lambda_{13}, \lambda_{14}) \\
&= \left(-\frac{\partial}{\partial\lambda_1}\right)^{i+1}\left(-\frac{\partial}{\partial\lambda_2}\right)^{j+1} \left(-\frac{\partial}{\partial\lambda_3}\right)^{k+1} \left(-\frac{\partial}{\partial \lambda_4}\right)^{l+1}  \\
&\kern 5pc \left(-\frac{\partial}{\partial\lambda_{12}}\right)^{m+1}\left(-\frac{\partial}{\partial\lambda_{13}}\right)^{n+1}\left(-\frac{\partial}{\partial\lambda_{14}}\right)^{p+1} \\
&\times I_{41}(-1,-1,-1, -1;-1,-1,-1; \lambda_1, \lambda_2, \lambda_3, \lambda_4 ; \lambda_{12},\lambda_{13}, \lambda_{14}) \cdot
\end{split}
\end{flalign}
The respective generating integral is given by 
\begin{flalign} %Eqn 52
\begin{split}
&I_{41}(-1, -1, -1, -1 ; -1, -1, -1 ; \lambda_1, \lambda_2, \lambda_3, \lambda_4 ; \lambda_{12},\lambda_{13}, \lambda_{14}) \\
&= \int {d\overrightarrow{r_1}d\overrightarrow{r_2}d\overrightarrow{r_3}d\overrightarrow{r_4} (r_1 r_2 r_3 r_4 r_{12}r_{13}r_{14})^{-1} } \\
&\times exp(-\lambda_1r_1-\lambda_2r_2-\lambda_3r_3-\lambda_4r_4-\lambda_{12}r_{12}-\lambda_{13}r_{13}-\lambda_{14}r_{14}),
\end{split}
\end{flalign}
and is denoted, in short, by $I_{41}^g(\lambda_1, \lambda_2, \lambda_3, \lambda_4 ; \lambda_{12},\lambda_{13}, \lambda_{14})$.
Similarly from  equation(50) we can express the generating integral as

\begin{flalign} %Eqn 53
\begin{split}
&I_{42}^g(\lambda_1, \lambda_2, \lambda_3, \lambda_4 ; \lambda_{21},\lambda_{23}, \lambda_{24})= \int {d\overrightarrow{r_1}d\overrightarrow{r_2}d\overrightarrow{r_3}d\overrightarrow{r_4} (r_1 r_2 r_3 r_4 r_{21}r_{23}r_{24})^{-1} }\\ 
&\kern 3pc \times exp(-\lambda_1r_1-\lambda_2r_2-\lambda_3r_3-\lambda_4r_4-\lambda_{21}r_{21}-\lambda_{23}r_{23}-\lambda_{24}r_{24})\cdot
\end{split}
\end{flalign}
The other two generating integrals corresponding to integrals $I_{43}$ and $I_{44}$ are, respectively, given by 

\begin{align} %Eqn54
\begin{split}
&I_{43}^g(\lambda_1, \lambda_2, \lambda_3, \lambda_4 ; \lambda_{31},\lambda_{32}, \lambda_{34})= \int {d\overrightarrow{r_1}d\overrightarrow{r_2}d\overrightarrow{r_3}d\overrightarrow{r_4} (r_1 r_2 r_3 r_4 r_{31}r_{32}r_{34})^{-1} }\\ 
&\kern 4pc \times exp(-\lambda_1r_1-\lambda_2r_2-\lambda_3r_3-\lambda_4r_4-\lambda_{31}r_{31}-\lambda_{32}r_{32}-\lambda_{34}r_{34}),
\end{split}
\end{align} %54
and 
\begin{align} %Eqn55
\begin{split}
I_{44}^g(\lambda_1, \lambda_2, \lambda_3, \lambda_4 ; \lambda_{41},\lambda_{42}, \lambda_{43})= \int {d\overrightarrow{r_1}d\overrightarrow{r_2}d\overrightarrow{r_3}d\overrightarrow{r_4} (r_1 r_2 r_3 r_4 r_{41}r_{42}r_{43})^{-1} }\\ \times exp(-\lambda_1r_1-\lambda_2r_2-\lambda_3r_3-\lambda_4r_4-\lambda_{41}r_{41}-\lambda_{42}r_{42}-\lambda_{43}r_{43})\cdot
\end{split}
\end{align} 

It is easy to observe that if any one of the generating integrals given by  equations(52-55) is evaluated analytically, the closed-form expressions for the other three can be written by inspection, keeping in mind that $r_{ij}=r_{ji}$ and $\lambda_{ij}=\lambda_{ji}$. Hence expression for all the general four - electron integrals $I_{41}, I_{42}, I_{43}$ and $I_{44}$ can be obtained by parametric differentiation method as mentioned in  equation(51). \\
\hfill \break
\textbf{5.1 (a)  Evaluation of the generating integral $I_{42}^g$} \\
\hfill \break 
Closed - form expression for the generating integral $I_{42}^g$ as defined in  equation(53) has already been reported in paper I. Only few steps will be repeated here before giving the final expression with the intention to obtain several limiting expressions for $I_{42}^g$ which were not reported earlier.
The integral given by  equation(53) can be recast as 
\begin{equation} % Equation 56
\begin{split}
&I_{42}^g(\lambda_1, \lambda_2, \lambda_3, \lambda_4 ; \lambda_{12},\lambda_{23}, \lambda_{24})= \int {d\overrightarrow{r_2}{r_2}^{-1} exp(-\lambda_2r_2) } \\
&\kern 6pc \times J(\lambda_3, \lambda_{23},r_2)J(\lambda_4, \lambda_{24}, r_2) J(\lambda_1, \lambda_{12}, r_2),
\end{split}
\end{equation}
where the J's are the integrals defined in  equation(1). Substituting proper closed-form expression for the J's from  equation(3) in  equation(56) and doing the angular integration, $I_{42}^g$ is reduced to the following one-dimensional form :
\begin{equation} %Equation 57
I_{42}^g =\int_0^\infty {dr_2 r_2^{-2} f(r_2)}, \\
\end{equation}
where the function A is given by 
\begin{equation}%Equation 58
A=256\pi^4[(\lambda_1^2 - \lambda_{12}^2)(\lambda_3^2 - \lambda_{23}^2)(\lambda_4^2 - \lambda_{24}^2)]^{-1} ,
\end{equation}
and the function $f(r_2)$ is a sum of eight terms of the form $exp(-\beta_ir_2), i=1,2,3,4, ....,8$. Each $\beta_i$ is a sum of four different $\lambda$'s out of the seven $\lambda$'s  in  equation(56), and the expressions for all the $\beta_i$'s are different.
 
It can be shown that all three functions $f(r_2)$ , $f'(r_2)$ and $f''(r_2)$ tend to zero as $r_2$ tends to zero. Here $f'(r_2)$ and $f''(r_2)$ represent the first order and the second order derivatives, respectively. Also by employing L'Hospital's rule for $0/0$, it is easy to prove that $\frac{f(r_2)}{r_2^2}\to 0, \frac{f(r_2)}{r_2}\to 0$ and $\frac{f'(r_2)}{r_2}\to 0$ as $r_2\to 0$. Integrating by parts, and then making use of  equation(19), the integral in  equation(57) is evaluated to obtain the following closed-form expression :
\begin{equation}%Equation 59
I_{42}^g=AL,
\end{equation}
where the function A is given by  equation(58) and the function L by
\begin{equation} %Equation 60
L(\lambda_1, \lambda_2,\lambda_3,\lambda_4 ; \lambda_{12}, \lambda_{23}, \lambda_{24})=\sum_{i=1}^8L_i,
\end{equation}
with
\begin{align*}
\begin{split}
&L_1=(\lambda_2+\lambda_{12}+\lambda_{23}) ln\left[\frac{(\lambda_2+\lambda_{12}+\lambda_{23}+
\lambda_{24})}{(\lambda_2+ \lambda_{12}+\lambda_{23}+\lambda_4)}\right],\\
%%%%%%%%%%%%
&L_2=(\lambda_2+\lambda_{12}+\lambda_3) ln\left[\frac{(\lambda_2+\lambda_{12}+\lambda_3+\lambda_4)}{(\lambda_2+ \lambda_{12}+\lambda_3+\lambda_{24})}\right], \\
%%%%%%%%%%%%%%%%%%
&L_3=(\lambda_1+\lambda_2+\lambda_{23}) ln\left[\frac{(\lambda_1+\lambda_2+\lambda_{23}+\lambda_4)}{(\lambda_1+ \lambda_2+\lambda_{23}+\lambda_{24})}\right], \\
%%%%%%%%%%%
&L_4=(\lambda_1+\lambda_2+\lambda_3) ln\left[\frac{(\lambda_1+\lambda_2+\lambda_3+\lambda_{24})}{(\lambda_1+ \lambda_2+\lambda_3+\lambda_4)}\right], \\
%%%%%%%%%%%
&L_5=\lambda_{24} ln\left[\frac{(\lambda_2+\lambda_{12}+\lambda_{23}+\lambda_{24})}{(\lambda_2+ \lambda_{12}+\lambda_3+\lambda_{24})}\right], \\
%%%%%%%%%%%%%%
&L_6=\lambda_{24} ln\left[\frac{(\lambda_1+\lambda_2+\lambda_3+\lambda_{24})}{(\lambda_1+ \lambda_2+\lambda_{23}+\lambda_{24})}\right],\\
%%%%%%%%%%%%%%
&L_7=\lambda_4 ln\left[\frac{(\lambda_2+\lambda_3+\lambda_4+\lambda_{12})}{(\lambda_2+ \lambda_4+\lambda_{12}+\lambda_{23})}\right], \\
%%%%%%%%%%%%
&L_8=\lambda_4 ln\left[\frac{(\lambda_1+\lambda_2+\lambda_4+\lambda_{23})}{(\lambda_1+ \lambda_2+\lambda_3+\lambda_4)}\right]. \\
%%%%%%%%%
\end{split}
\end{align*}

It is worth mentioning here that an alternative expression with larger symmetry than the one reported in paper I, and now given by  equation(59) above, for $I_{42}^g$, has been reported very recently by King [59]. However, by minor manipulations, it is easily shown that the above expression becomes exactly identical with that given by  equation(35) in [59], and hence, all the following discussions are made relating to the expression given by  equation(59).\\
\hfill \break
\textbf{Limiting expressions for $I_{42}^g$} \\
\hfill \break
If, as a special case, $\lambda_{12}=\lambda_{23}=\lambda_{24}=0$, then, after some manipulations  equation(59) simplifies to 
\begin{equation}
\begin{split} %Equation 61
&I_{42}^g(\lambda_1, \lambda_2, \lambda_3, \lambda_4 ; 0,0,0)= 256\pi^4(\lambda_1\lambda_3\lambda_4)^{-2}[\lambda_2ln\lambda_2-(\lambda_2+\lambda_1)ln(\lambda_2+\lambda_1)\\
&-(\lambda_2+\lambda_3)ln(\lambda_2+\lambda_3)-(\lambda_2+\lambda_4)ln(\lambda_2+\lambda_4)+(\lambda_2+\lambda_3+\lambda_4)ln(\lambda_2+\lambda_3+\lambda_4)\\&+(\lambda_2+\lambda_1+\lambda_4)ln(\lambda_2+\lambda_1+\lambda_4)+(\lambda_2+\lambda_1+\lambda_3)ln(\lambda_2+\lambda_1+\lambda_3)\\
& -(\lambda_1+\lambda_2+\lambda_3+\lambda_4)ln(\lambda_1+\lambda_2+\lambda_3+\lambda_4)\big]\cdot
\end{split}
\end{equation} %61

It is worth pointing out here that a closed-form expression for the integral on the left hand side of  equation(61) was obtained by King [52] directly, by employing expansion formula of Sack [71] and of Perkins [72] for $r_{ij}^\nu$.  The corresponding expression contains a minor typographical error which is corrected by changing the signs before the fourth and the fifth terms within the curly brackets on the right hand side of  equation(44) of the reported paper [52], which has been pointed out very recently in [59]. Incorporating these minor corrections, it is easy to show that the corrected expression becomes identical with the one given on the right hand side of  equation(61) obtained, as a special case, from the generating integral $I_{42}^g$ defined in  equation(53) and evaluated and reported earlier in paper I.

By differentiating both side of  equation(59) with respect to $\lambda_2$, a closed-form expression for the integral $I_{42}(-1, 0, -1, -1,; -1, -1, -1, ;\lambda_1, \lambda_2, \lambda_3, \lambda_4; \lambda_{12}, \lambda_{23}, \lambda_{24})$, as a special case, is obtained which is exactly identical with the right hand side expression in  equation(32) of the recent paper [59], wherein the integral has been evaluated directly. Then substituting $\lambda_{12}=\lambda_{23}=\lambda_{24}=0$ in that expression the following integral is evaluated :
\begin{multline} %Eqn62
I_{42}^g (-1,0,-1,-1;-1,-1,-1;\lambda_1\lambda_2,\lambda_3,\lambda_4 ; 0,0,0)\\
=\frac{256\pi^4}{(\lambda_1\lambda_3\lambda_4)^2} ln\left[\frac{(\lambda_2+\lambda_1)(\lambda_2+\lambda_3)(\lambda_2+\lambda_4)(\lambda_1+\lambda_2+\lambda_3+\lambda_4)}{\lambda_2(\lambda_2+\lambda_3+\lambda_4)(\lambda_2+\lambda_1+\lambda_4)(\lambda_2+\lambda_1+\lambda_3)}\right] ,
\end{multline}
which is in conformity with  equation(16) of the report of Roberts [65] who had evaluated the integral by expanding $(r_{12}r_{23}r_{24})^{-1}$ in spherical harmonics. Also it is easy to establish  equation(62) by differentiating both sides of  equation(61) with respect to  $\lambda_2$.  Further, if $\lambda_1=\lambda_2=\lambda_3=\lambda_4=\delta$, then  equation(61) becomes

\begin{equation}%63
I_{42}^g(\delta,\delta,\delta,\delta;0,0,0)=256\pi^4\delta^{-5}(9\: ln3 - 14\: ln2)
\end{equation}
and  equation(62) becomes
\begin{equation}%64
I_{42}(-1,0,-1,-1;-1,-1,-1;\delta,\delta,\delta,\delta;0,0,0)=256\pi^4\delta^{-6}(5\:ln2 - 3\:ln3)\cdot
\end{equation}
The expressions obtained in  equations(62-64) here have also been reported in [52].

It is observed in  equation(60) that if $\lambda_{12}\rightarrow\lambda_1$,  then $L\rightarrow0$. Similarly if $\lambda_{23}\rightarrow\lambda_3$, then also $L\rightarrow0$. So also $L\rightarrow0$ if $\lambda_{24}\rightarrow\lambda_4$. In such cases the right hand side expression in  equation(59) assumes $0/0$ form. Hence, finite expression for $I_{42}^g$ in various limiting cases can be obtained by applying L'Hospital's rule for $0/0$.

Thus if $\lambda_{12}\to \lambda_1$  alone,  equation(59) gives
\begin{multline} %Eqn65
I_{42}^g (\lambda_1,\lambda_2,\lambda_3,\lambda_4 ; \lambda_1, \lambda_{23}, \lambda_{24})\\
=\frac{128\pi^4}{\lambda_1(\lambda_3^2-\lambda_{23}^2)(\lambda_4^2-\lambda_{24}^2)} ln\left[\frac{\lambda_1+\lambda_2+\lambda_{23}+\lambda_4}{\lambda_1+\lambda_2+\lambda_{23}+\lambda_{24}} \times \frac{\lambda_1+\lambda_2+\lambda_3+\lambda_{24}}{\lambda_1+\lambda_2+\lambda_3+\lambda_4}\right]\cdot
\end{multline}
If $\lambda_{23}\rightarrow\lambda_3$ alone in  equation(59), the result is 
\begin{multline} %Eqn66
I_{42}^g (\lambda_1,\lambda_2,\lambda_3,\lambda_4 ; \lambda_{12}, \lambda_3, \lambda_{24})\\
=\frac{128\pi^4}{(\lambda_1^2-\lambda_{12}^2)\lambda_3(\lambda_4^2-\lambda_{24}^2)} ln\left[\frac{\lambda_2+\lambda_3+\lambda_{12}+\lambda_4}{\lambda_2+\lambda_3+\lambda_{12}+\lambda_{24}} \times \frac{\lambda_2+\lambda_3+\lambda_1+\lambda_{24}}{\lambda_2+\lambda_3+\lambda_1+\lambda_4}\right] \cdot
\end{multline}
On the other hand, if $\lambda_{24}\rightarrow\lambda_4$ alone,  equation(59) reduces to 
\begin{multline} %Eqn67
I_{42}^g (\lambda_1,\lambda_2,\lambda_3,\lambda_4 ; \lambda_{12}, \lambda_{23}, \lambda_4)\\
=\frac{128\pi^4}{(\lambda_1^2-\lambda_{12}^2)(\lambda_3^2-\lambda_{23}^2)\lambda_4} ln\left[\frac{\lambda_2+\lambda_{12}+\lambda_3+\lambda_4}{\lambda_2+\lambda_{12}+\lambda_{23}+\lambda_4} \times \frac{\lambda_1+\lambda_2+\lambda_{23}+\lambda_4}{\lambda_1+\lambda_2+\lambda_3+\lambda_4}\right]\cdot
\end{multline}

In case $\lambda_{12}\to \lambda_1$ and $\lambda_{23}\to \lambda_3$ together, applying L'Hospital's rule for $0/0$ to the right hand side expressions in  equation(65) or  equation(66) as required, the following result is obtained :
\begin{multline} %Eqn68
I_{42}^g (\lambda_1,\lambda_2,\lambda_3,\lambda_4 ; \lambda_1, \lambda_3, \lambda_{24})\\
=64\pi^4[\lambda_1\lambda_3(\lambda_4+\lambda_{24})(\lambda_1+\lambda_2+\lambda_3+\lambda_4)(\lambda_1+\lambda_2+\lambda_3+\lambda_{24})]^{-1}\cdot
\end{multline}
If, further, $\lambda_{24}=\lambda_4$, then
\begin{equation} %Eqn69
I_{42}^g (\lambda_1,\lambda_2,\lambda_3,\lambda_4 ; \lambda_1, \lambda_3, \lambda_4)=32\pi^4[\lambda_1\lambda_3\lambda_4(\lambda_1+\lambda_2+\lambda_3+\lambda_4)^2]^{-1}\cdot
\end{equation}

In the most special case, if each of the exponential parameters( $\lambda$'s$)$ in  equation(59) is equal to $\delta$,  then  equation(69) simplifies to 

\begin{equation} %Eqn70
I_{42}^g (\delta,\delta, \delta, \delta; \delta, \delta, \delta)=2\pi^4\delta^{-5}\cdot
\end{equation}

Adopting the same procedure as above, expressions for the following integrals are obtained from  equation(67) taking limits $\lambda_{12}\to \lambda_1$ and $\lambda_{23}\to \lambda_3$, respectively :
\begin{equation}%Equation 71
\begin{split}
&I_{42}^g (\lambda_1,\lambda_2,\lambda_3,\lambda_4 ; \lambda_1, \lambda_{23}, \lambda_4)\\
&=64\pi^4[\lambda_1(\lambda_3+\lambda_{23})\lambda_4(\lambda_2+\lambda_1+\lambda_3+\lambda_4)(\lambda_2+\lambda_1+\lambda_{23}+\lambda_4)]^{-1},
\end{split}
\end{equation}
\begin{equation}%Equation 72
\begin{split}
&I_{42}^g (\lambda_1,\lambda_2,\lambda_3,\lambda_4 ; \lambda_{12}, \lambda_3, \lambda_4)\\
&=64\pi^4[(\lambda_1+\lambda_{12})\lambda_3\lambda_4(\lambda_1+\lambda_2+\lambda_3+\lambda_4)(\lambda_2+\lambda_{12}+\lambda_3+\lambda_4)]^{-1}\cdot
\end{split}
\end{equation}
As expected,  equation(69) is again established if either $\lambda_{23}$ is replaced by $\lambda_3$ in  equation(71) or $\lambda_{12}$ is replaced by $\lambda_1$ in  equation(72).

It is worth pointing out here that all the limiting expressions for $I_{42}^g$ given in  equations(61) and (65-72) can also be obtained directly from  equation(53) by making use of  equation(3) and/or  equation(4) in the integral in  equation(56) as per the desired limits and carrying out the integration. This statement has been actually verified by evaluating the integrals.\\
\hfill\break
\textbf{5.1(b)   Evaluation of the generating integral $I_{41}^g$} \\ 
\hfill\break
The generating integral $I_{41}^g$ as given by  equation(52) can be recast as 
\begin{equation}%Equation 73
\begin{split}
&I_{41}^g (\lambda_1,\lambda_2,\lambda_3,\lambda_4 ; \lambda_{12}, \lambda_{13}, \lambda_{14})\\
&=\int{d\overrightarrow{r_1}[r_1^{-1} exp(-\lambda_1r_1)]}J(\lambda_2,\lambda_{12},r_1) J(\lambda_3,\lambda_{13},r_1) J(\lambda_4, \lambda_{14},r_1),
\end{split}
\end{equation}
where the RJ's are the integrals given by  equations(1) and (3). Performing the evaluation as outlined in the case of the generating integral $I_{42}^g$, the following closed-form expression is obtained for $I_{41}^g$ :
\begin{align} %Eq 74
&I_{41}^g (\lambda_1,\lambda_2,\lambda_3,\lambda_4 ; \lambda_{12}, \lambda_{13}, \lambda_{14}) = BM,
\end{align}
where $B$ and $M$ are functions given by 
\begin{equation} %Eqn75
B=256\pi^4[(\lambda_2^2-\lambda_{12}^2)(\lambda_3^2-\lambda_{13}^2)(\lambda_4^2-\lambda_{14}^2)]^{-1},
\end{equation}
and
\begin{equation}%Eqn76
M(\lambda_1,\lambda_2, \lambda_3, \lambda_4 ; \lambda_{12}, \lambda_{13}, \lambda_{14})=\sum_{i=1}^8M_i ,
\end{equation}
with
\begin{align*}
\begin{split}
&M_1=(\lambda_1+\lambda_{12}+\lambda_{13})ln\bigg[\frac{(\lambda_1+\lambda_{12}+\lambda_{13}+\lambda_{14})}{(\lambda_1+\lambda_{12}+\lambda_{13}+\lambda_4)}\bigg] ,\\
%%%%%%%%%%%%%%
&M_2=(\lambda_1+\lambda_{12}+\lambda_3)ln\bigg[\frac{(\lambda_1+\lambda_{12}+\lambda_3+\lambda_4)}{(\lambda_1+\lambda_{12}+\lambda_3+\lambda_{14})}\bigg] ,\\
%%%%%%%%%%%
&M_3=(\lambda_2+\lambda_1+\lambda_{13})ln\bigg[\frac{(\lambda_2+\lambda_1+\lambda_{13}+\lambda_4)}{(\lambda_2+\lambda_1+\lambda_{13}+\lambda_{14})}\bigg] ,\\
%%%%%%%%%%
&M_4=(\lambda_2+\lambda_1+\lambda_3)ln\bigg[\frac{(\lambda_2+\lambda_1+\lambda_3+\lambda_{14})}{(\lambda_2+\lambda_1+\lambda_3+\lambda_4)}\bigg] ,\\
%%%%%%%%%%%
&M_5=\lambda_{14}ln\bigg[\frac{(\lambda_1+\lambda_{12}+\lambda_{13}+\lambda_{14})}{(\lambda_1+\lambda_{12}+\lambda_3+\lambda_{14})}\bigg] ,\\
%%%%%%%%%%%
&M_6=\lambda_{14}ln\bigg[\frac{(\lambda_2+\lambda_1+\lambda_3+\lambda_{14})}{(\lambda_2+\lambda_1+\lambda_{13}+\lambda_{14})}\bigg] ,\\
%%%%%%%%%%
&M_7=\lambda_4ln\bigg[\frac{(\lambda_1+\lambda_3+\lambda_4+\lambda_{12})}{(\lambda_1+\lambda_4+\lambda_{12}+\lambda_{13})}\bigg],\\
%%%%%%%%%%
&M_8=\lambda_4ln\bigg[\frac{(\lambda_2+\lambda_1+\lambda_4+\lambda_{13})}{(\lambda_2+\lambda_1+\lambda_3+\lambda_4)}\bigg] \cdot
\end{split}
\end{align*}

Noting that $r_{ij}=r_{ji}$  and  $\lambda_{ij} = \lambda_{ji}$,  equation(52) for $I_{41}^g$ is obtained from  equation(53) for $I_{42}^g$ by making the interchange $(2\leftrightarrows1)$ in  equation(53). Comparing the expression for $I_{42}^g$ given in the right hand side of  equation(59) with that for $I_{41}^g$ given in  equation(74) , it is clearly observed that the right hand side expression for $I_{41}^g$  in  equation(74) is exactly obtained from the right hand side expression for $I_{42}^g$ in  equation(59) by performing the interchange $(2\leftrightarrows1)$ in  equation(59), as pointed out earlier. Thus we conclude that even without going through the process of evaluation, the closed-form expression for $I_{41}^g$ can be obtained by inspection from that of $I_{42}^g$ .\\
\hfill\break
\textbf{Limiting expressions for $I_{41}^g$}\\
\hfill\break
If, as a special case, $\lambda_{12}=\lambda_{13}=\lambda_{14}=0$ substituted on both sides of  equation(74), an expression for the integral  $I_{41}^g(\lambda_1, \lambda_2, \lambda_3, \lambda_4; 0, 0, 0)$ is obtained which is observed to be exactly identical with the one arrived at from the right hand side of  equation(61) by making the interchange $(2\leftrightarrows1)$, as expected. Thus the following equality is obtained :

\begin{equation} %Equation 77
I_{41}^g(\lambda_1, \lambda_2, \lambda_3, \lambda_4; 0, 0, 0)=I_{42}^g(\lambda_1\leftrightarrows\lambda_2, \lambda_3, \lambda_4; 0, 0, 0) \cdot
\end{equation}
Various other limiting expressions for $I_{41}^g$, similar to  equations(65-72) in case of $I_{42}^g$, can be easily obtained following the procedure adopted for $I_{42}^g$. \\
\hfill\break
\textbf{Expression for $<r_{12} r_{13} / r_{14}>$ }\\
\hfill\break
Sometimes it is required to obtain the expectation value $<r_{12} r_{13} / r_{14}>$ in linear theories of atoms with four or more number of electrons. This can be achieved by differentiating both sides of  equation(74) twice with respect to $\lambda_{12}$ and twice with respect to $\lambda_{13}$. Thus the required integral is 
\begin{equation} %Equation 78
I_{41}(-1, -1, -1, -1; 1, 1, -1 ; \lambda_1, \lambda_2, \lambda_3, \lambda_4 ; \lambda_{12}, \lambda_{13}, \lambda_{14}) = \frac{\partial^2}{\partial\lambda_{12}^2}\frac{\partial^2}{\partial\lambda_{13}^2}[BM]\cdot
\end{equation}

Carrying out the differentiations on the right hand side, an analytical expression is obtained for the integral on the left hand side of  equation(78), which is related to $<r_{12} r_{13}/r_{14}>$.

If $\lambda_{12}=\lambda_{13}=\lambda_{14}=0$, as a special case, then the following expression is obtained :

\begin{flalign} %Equation 79
\begin{split}
&I_{41}(-1, -1, -1, -1 ; 1, 1, -1 ; \lambda_1, \lambda_2, \lambda_3, \lambda_4 ; 0, 0, 0,) \\
&=\frac{4}{\lambda_2^2\lambda_3^2}I_{41}^g(\lambda_1, \lambda_2, \lambda_3, \lambda_4 ; 0,0,0) \\
&+\frac{512\pi^4}{\lambda_2^2\lambda_4^2} \Bigg[\frac{1}{\lambda_2^2\lambda_3^2}\left(\frac{1}{\lambda_1}-\frac{1}{\lambda_1+\lambda_2}-\frac{1}{\lambda_1+\lambda_4}+\frac{1}{\lambda_1+\lambda_2+\lambda_4}\right) +\frac{1}{\lambda_3^2\lambda_1^3}\\
&-\frac{1}{\lambda_3^2(\lambda_1+\lambda_4)^3} +\frac{1}{\lambda_3^4}\left(\frac{1}{\lambda_1}-\frac{1}{\lambda_1+\lambda_3}-\frac{1}{\lambda_1+\lambda_4}+\frac{1}{\lambda_1+\lambda_3+\lambda_4}\right) \Bigg] \cdot
\end{split}
\end{flalign}
If, further, $\lambda_1=\lambda_2=\lambda_3=\lambda_4=\delta$, then, the above equation simplifies to
\begin{equation} %Equation 80
\begin{split}
I_{41}(-1, -1, -1, -1 ; 1, 1, -1 ; \delta, \delta, \delta, \delta; 0, 0, 0,) \\
=512\pi^4\delta^{-9}(37/24 +18\:ln3-28\:ln2)\cdot
\end{split}
\end{equation}

Following the same procedure as adopted in  equation(78), the expectation values $\left<r_{12}r_{14}/r_{13}\right>$ and $\left<r_{13}r_{14}/r_{12}\right>$ can be obtained. \\ 
\hfill\break
\textbf{5.1 (c) Evaluation of the generating integrals $I_{43}^g$ and $I_{44}^g$}\\
\hfill\break
Closed-form expression for $I_{43}^g$ as defined in  equation(54) can be obtained from that of $I_{42}^g$ as given in the right hand side of  equation(59) by performing the interchange  $(2\leftrightarrows3)$  in  equation(59). Alternatively, the integral $I_{43}^g$ can be evaluated directly as was done in the case of $I_{42}^g$ employing  equations(1) and (3). It has been observed that the two expressions obtained for $I_{43}^g$ by the above two alternative approaches do not appear to be identical. However, by minor manipulative algebra, both the expressions are shown to be exactly identical. In yet another alternative approach, expression for $I_{43}^g $ can be obtained from the right hand side expression in  equation(74) for $I_{41}^g$, by making the interchange $(1\leftrightarrows3)$ in  equation(74).

Expression for $I_{44}^g$ as defined in  equation(55) can be derived directly employing  equations(1) and (3) as was done for $I_{42}^g$. Alternatively, by inspection, its expression can be written as the one obtained either from  equation(59) by the interchange $(2\leftrightarrows4)$, or from  equation(74) by the interchange $(1\leftrightarrows4)$.

It is concluded, in general, that closed-form expression for  $I_{4i}^g$ can be obtained from the expression for $I_{4j}^g$, by making interchange  $(j \leftrightarrows i)$ in the expression for $I_{4j}^g$, and vice versa, as pointed out earlier.

Various limiting expressions for $I_{43}^g$ and $I_{44}^g$ can be derived by following the same procedure adopted for $I_{42}^g$ while obtaining expressions in  equations(61) and (65-72).
\\
\hfill\break
\textbf{5.2 Integrals represented by open square diagrams}
\par \textbf{I.} There is one set of four general integrals involving exponential correlation corresponding to four equivalent open square diagrams, with one side of each square missing. Thus the respective integrals involve correlations of the form $(r_{ij}r_{jk}r_{kl})^{-1}exp(-\lambda_{ij}r_{ij}-\lambda_{jk}r_{jk}-\lambda_{kl}r_{kl})$  with i,j,k and l cyclic, beginning with i=1. For example, for the first generating integral i=1, j=2, k=3 and l=4; for the second i=2, j=3, k=4 and l=1, etc.

\textbf{II.} There is another set of four such general four-electron integrals, each represented by a open square diagram with two opposite sides of the square missing and the other two opposite sides connected by a diagonal. As an example, the correlation of the form $(r_{12}r_{13}r_{34})^{-1}exp(-\lambda_{12}r_{12}-\lambda_{13}r_{13}-\lambda_{34}r_{34})$ is considered corresponding to an open square with opposite sides represented by $r_{12}$ and $r_{34}$, respectively, and connected by the diagonal respresented by $r_{13}$. The other correlations considered are of the form $(r_{12} r_{24} r_{43})^{-1} exp (-\lambda_{12}r_{12}-\lambda_{24}r_{24}-\lambda_{43}r_{43})$,  $(r_{23} r_{31} r_{14})^{-1} exp (-\lambda_{23}r_{23}-\lambda_{31}r_{31}-\lambda_{14}r_{14})$, and $(r_{23}r_{24}r_{41})^{-1}exp(-\lambda_{23}r_{23}-\lambda_{24}r_{24}-\lambda_{41}r_{41})$.
\\
\hfill\break
\textbf{5.2 (I) Integrals belonging to category (I)} 

The four general integrals belonging to the first category are denoted as (a) $K_{41}$, (b) $K_{42}$, (c) $K_{43}$ and (d) $K_{44}$, and defined as 
\begin{equation} %Equation 81
\begin{split}
&K_{41}(i, j ,k , l; m , n, p ; \lambda_1, \lambda_2, \lambda_3, \lambda_4 ; \lambda_{12}, \lambda_{23}, \lambda_{34}) \\
&=\int{d\overrightarrow{r_1}d\overrightarrow{r_2}d\overrightarrow{r_3}d\overrightarrow{r_4}r_1^i r_2^j r_3^k r_4^l r_{12}^m r_{23}^n r_{34}^p} \\
&\times exp(-\lambda_1r_1 - \lambda_2r_2 -\lambda_3r_3 - \lambda_4r_4-\lambda_{12}r_{12}-\lambda_{23}r_{23} - \lambda_{34}r_{34}),
\end{split}
\end{equation}
with similar definitions for $K_{42}, K_{43}$ and $K_{44}$.

The integral $K_{41}$ can be related to the corresponding generating integral $K_{41}^g$, as usual, through parametric differentiation. Thus
\begin{align}%Equation 82
\begin{split}
&K_{41}(i, j, k, l ; m, n, p ; \lambda_1, \lambda_2, \lambda_3, \lambda_4 ; \lambda_{12}, \lambda_{23}, \lambda_{34}) \\
&=\left(-\frac{\partial}{\partial\lambda_1}\right)^{i+1}\left(-\frac{\partial}{\partial\lambda_2}\right)^{j+1} \left(-\frac{\partial}{\partial\lambda_3}\right)^{k+1}\left(-\frac{\partial}{\partial\lambda_4}\right)^{l+1} \\& \left(-\frac{\partial}{\partial\lambda_{12}}\right)^{m+1}\left(-\frac{\partial}{\partial\lambda_{23}}\right)^{n+1} \left(-\frac{\partial}{\partial\lambda_{34}}\right)^{p+1} K_{41}^g,
\end{split}
\end{align}
where $K_{41}^g$ is the generating integral given by 

\begin{align}%Equation 83
\begin{split}
&K_{41}^g(\lambda_1, \lambda_2, \lambda_3, \lambda_4 ; \lambda_{12}, \lambda_{23}, \lambda_{34}) = \int{d\overrightarrow{r_1}d\overrightarrow{r_2}d\overrightarrow{r_3}d\overrightarrow{r_4} (r_1r_2r_3r_4r_{12}r_{23} r_{34})^{-1}} \\
&\kern 3pc \times exp(-\lambda_1r_1 -\lambda_2r_2 -\lambda_3r_3 -\lambda_4r_4 -\lambda_{12}r_{12} -\lambda_{23}r_{23}- \lambda_{34}r_{34}),
\end{split}
\end{align}
with similar definitions for the other three integrals and the respective generating integrals.

\textbf{(a)  Analytic evaluation of $K_{41}^g$ }

The integral $K_{41}^g$ as defined in  equation(83) can be written as
\begin{align}%Equation 84
\begin{split}
&K_{41}^g= \int{d\overrightarrow{r_2}d\overrightarrow{r_3}(r_2r_3r_{23})^{-1}} exp( -\lambda_2r_2 -\lambda_3r_3 -\lambda_{23}r_{23}) \\
&\kern 7pc \times J(\lambda_1, \lambda_{12}, r_2) J(\lambda_4, \lambda_{34}, r_3),
\end{split}
\end{align}
where the J integrals are defined by  equation(1) with their closed-form expressions given by  equation(3). Substituting the expressions for the J's in  equation(84), it simplifies to 
\begin{align}%Equation 85
\begin{split}
&K_{41}^g= D[I_2(-2, -2,-1;\alpha_1, \beta_1, \gamma_1)-I_2(-2, -2,-1;\alpha_2, \beta_2, \gamma_2) \\
&\kern 3pc -I_2(-2, -2,-1;\alpha_3, \beta_3, \gamma_3)+ I_2(-2, -2,-1;\alpha_4, \beta_4, \gamma_4)],
\end{split}
\end{align}
where D is a function given by
\begin{equation}%Equation 86
D(\lambda_1, \lambda_{12}, \lambda_4, \lambda_{34})=16\pi^2[(\lambda_1^2-\lambda_{12}^2)(\lambda_4^2-\lambda_{34}^2)]^{-1},
\end{equation}
and, in general,
\begin{equation}%Equation 87
I_2(-2, -2, -1; \alpha, \beta, \gamma)=\int{d\overrightarrow{r_2}d\overrightarrow{r_3}(r_2^2r_3^2r_{23})^{-1}} \times exp(-\alpha r_2-\beta r_3 -\gamma r_{23}),
\end{equation}
with $\alpha_1 = \alpha_2 = \lambda_2 + \lambda_{12}$ , $\alpha_3 = \alpha_4 = \lambda_1 + \lambda_2$, $\beta_1 = \beta_3 = \lambda_3 + \lambda_{34}$, $\beta_2 = \beta_4 = \lambda_3 + \lambda_4$ and $\gamma_1 = \gamma_2 = \gamma_3 = \gamma_4 = \lambda_{23}$.

The analytical expression for the integral defined in  equation(87), in general, has been given in  equation(25). Replacing the $I_2$ integrals in  equation(85) by their closed-form expressions, and doing some simplifications, the following expression for $K_{41}^g$ is successfully obtained :

\begin{equation}%Equation 88
K_{41}^g(\lambda_1, \lambda_2, \lambda_3, \lambda_4, \lambda_{12}, \lambda_{23}, \lambda_{34}) = DN, 
\end{equation}
where the function D is given by  equation(86) and the function N, which is the sum of the four $I_2$ integrals within the square brackets on the right hand side of  equation(85), is given by 

\begin{align}%Equation 89
 \begin{split}
&N(\lambda_1, \lambda_2, \lambda_3, \lambda_4; \lambda_{12}, \lambda_{23}, \lambda_{34}) = \frac{8\pi^2}{\lambda_{23}}\bigg[ ln \left( \frac{\lambda_{12}+\lambda_2+\lambda_{23}}{\lambda_1+\lambda_2+\lambda_{23}}\right) ln \left( \frac{\lambda_3+\lambda_4+\lambda_{23}}{\lambda_3+\lambda_{34}+\lambda_{23}}\right) \\
&\kern 6pc + dilog \left( \frac{\lambda_2+\lambda_{12}+\lambda_3+\lambda_{34}}{\lambda_2+\lambda_{12}+\lambda_{23}}\right) + dilog \left( \frac{\lambda_2+\lambda_{12}+\lambda_3+\lambda_{34}}{\lambda_3+\lambda_{34}+\lambda_{23}}\right) \\
&\kern 6pc - dilog \left( \frac{\lambda_2+\lambda_{12}+\lambda_3+\lambda_4}{\lambda_2+\lambda_{12}+\lambda_{23}}\right) - dilog \left( \frac{\lambda_2+\lambda_{12}+\lambda_3+\lambda_4}{\lambda_3+\lambda_4+\lambda_{23}}\right) \\
&\kern 6pc - dilog \left( \frac{\lambda_1+\lambda_2+\lambda_3+\lambda_{34}}{\lambda_1+\lambda_2+\lambda_{23}}\right) - dilog \left( \frac{\lambda_1+\lambda_2+\lambda_3+\lambda_{34}}{\lambda_3+\lambda_{34}+\lambda_{23}}\right) \\
& \kern 6pc + dilog \left( \frac{\lambda_1+\lambda_2+\lambda_3+\lambda_4}{\lambda_1+\lambda_2+\lambda_{23}}\right) + dilog \left( \frac{\lambda_1+\lambda_2+\lambda_3+\lambda_{34}}{\lambda_3+\lambda_4+\lambda_{23}}\right)\bigg].
\end{split}
\end{align}
It is worth pointing out here that an alternative analytical expression for the above integral has also been reported in [59] very recently. However, here, the following discussions are made relating to the right hand side expressions in  equations(88) and (85) only.

\textbf{Limiting expressions for $K_{41}^g$}

If $\lambda_{23}\rightarrow0$, the proper expressions for the $I_2$ integrals as per  equation(29) are substituted in  equation(85) to obtain, as a special case,
\begin{align}%Equation 90
\begin{split}
&K_{41}^g(\lambda_1, \lambda_2, \lambda_3, \lambda_4 ; \lambda_{12}, 0, \lambda_{34}) \\
&\kern 5pc =256\pi^4[(\lambda_1^2-\lambda_{12}^2)(\lambda_4^2-\lambda_{34}^2)]^{-1}[T_1 +T_2+T_3+T_4],
\end{split}
\end{align}
where
\begin{align*}
\begin{split}
&T_1 = \frac{1}{\lambda_1+\lambda_2} ln\left(\frac{\lambda_1+\lambda_2+\lambda_3+\lambda_4}{\lambda_3+\lambda_4}\times \frac{\lambda_3+\lambda_{34}}{\lambda_1+\lambda_2+\lambda_3+\lambda_{34}}\right), \\
&T_2 = \frac{1}{\lambda_3+\lambda_4} ln\left(\frac{\lambda_1+\lambda_2+\lambda_3+\lambda_4}{\lambda_1+\lambda_2}\times \frac{\lambda_2+\lambda_{12}}{\lambda_2+\lambda_{12}+\lambda_3+\lambda_4}\right), \\
&T_3 = \frac{1}{\lambda_2+\lambda_{12}} ln\left(\frac{\lambda_2+\lambda_{12}+\lambda_3+\lambda_{34}}{\lambda_3+\lambda_{34}}\times \frac{\lambda_3+\lambda_4}{\lambda_2+\lambda_{12}+\lambda_3+\lambda_4}\right), \\
&T_4 = \frac{1}{\lambda_3+\lambda_{34}} ln\left(\frac{\lambda_2+\lambda_{12}+\lambda_3+\lambda_{34}}{\lambda_2+\lambda_{12}}\times \frac{\lambda_1+\lambda_2}{\lambda_1+\lambda_2+\lambda_3+\lambda_{34}}\right), \\
\end{split}
\end{align*}
which is exactly identical with  equation(41) of the very recent paper [ ], wherein the integral on the left hand side of  equation(90) has been evaluated directly. Expressions for the integrals $K_{41}^g(\lambda_1, \lambda_2, \lambda_3, \lambda_4 ; 0, 0 , \lambda_{34}) $ and $K_{41}^g(\lambda_1, \lambda_2, \lambda_3, \lambda_4 ; \lambda_{12}, 0 , 0) $ can be easily obtained from  equation(90) by taking $\lambda_{12}=0$ and $\lambda_{34}=0$, respectively, on both sides. If both $\lambda_{12}=\lambda_{34}=0$, then  equation(90) simplifies to give
\begin{align}%Equation 91
\begin{split}
&K_{41}^g(\lambda_1, \lambda_2, \lambda_3, \lambda_4 ; 0 , 0, 0) \\
&= \frac{256\pi^4}{\lambda_1^2\lambda_4^2} \Bigg[ \frac{1}{\lambda_1+\lambda_2} ln\left(\frac{\lambda_1+\lambda_2+\lambda_3+\lambda_4}{\lambda_3+\lambda_4} 
\times \frac{\lambda_3}{\lambda_1 + \lambda_2 + \lambda_3}\right) \\
&\kern 7pc + \frac{1}{\lambda_3+\lambda_4} ln\left(\frac{\lambda_1+\lambda_2+\lambda_3+\lambda_4}{\lambda_1+\lambda_2} 
\times \frac{\lambda_2}{\lambda_2 + \lambda_3 + \lambda_4}\right) \\
& + \frac{1}{\lambda_2} ln\left(\frac{\lambda_2+\lambda_3}{\lambda_3} 
\times \frac{\lambda_3+\lambda_4}{\lambda_2 + \lambda_3 + \lambda_4}\right)+\frac{1}{\lambda_3} ln\left(\frac{\lambda_2+\lambda_3}{\lambda_2} 
\times \frac{\lambda_1+\lambda_2}{\lambda_1 + \lambda_2 + \lambda_3}\right) \Bigg],
\end{split}
\end{align}
which is exactly identical with  equation(40) of [52] and  equation(42) of [59]. If, further, $\lambda_1=\lambda_2=\lambda_3=\lambda_4=\delta , $ then
\begin{equation}%Equation 92
K_{41}^g(\delta, \delta, \delta, \delta ; 0 , 0, 0) =256 \pi^4(5~ln2-3~ln3)\times \delta^{-5}, 
\end{equation}
which is identical with  equation(41) of [52]. It is also observed in  equation(90) that if $\lambda_{12} \to  \lambda_1$, the right hand side expression assumes $0/0$ form. Employing L'Hospital's rule for $0/0$, it can be shown that
\begin{align}%Equation 93
\begin{split}
&K_{41}^g(\lambda_1, \lambda_2, \lambda_3, \lambda_4; \lambda_1, 0, \lambda_{34}) \\
& = \frac{128\pi^4}{\lambda_1(\lambda_1+\lambda_2)^2(\lambda_4^2-\lambda_{34}^2)} ln \left(\frac{\lambda_1+\lambda_2+\lambda_3+\lambda_{34}}{\lambda_1+\lambda_2+\lambda_3+\lambda_4} \times \frac{\lambda_3+\lambda_4}{\lambda_3+\lambda_{34}}\right)\cdot
\end{split}
\end{align}

If, further, $\lambda_{34} \to \lambda_4$, employing L'Hospital's rule for $0/0$,  equation(93) simplifies to give
\begin{align}%Equation 94
\begin{split}
&K_{41}^g(\lambda_1, \lambda_2, \lambda_3, \lambda_4; \lambda_1, 0, \lambda_4) \\
&= 64\pi^4\left[\lambda_1\lambda_4(\lambda_1+\lambda_2)(\lambda_3+\lambda_4)(\lambda_1+\lambda_2+\lambda_3+\lambda_4)\right]^{-1}\cdot
 \end{split}
\end{align}

In the most special case, if $\lambda_1=\lambda_2=\lambda_3=\lambda_4=\delta$,  then 

\begin{equation}%Equation 95
K_{41}^g(\delta, \delta, \delta, \delta ; \delta , 0, \delta) = 4\pi^4\delta^{-5} \cdot
\end{equation}
If $\lambda_{34}\rightarrow\lambda_4$ in  equation(90), employing L'Hospital's rule for $0/0$ , the following integral is evaluated :

\begin{align}  %Equation-96
\begin{split}
&K_{41}^g(\lambda_1, \lambda_2, \lambda_3, \lambda_4; \lambda_{12}, 0, \lambda_4) \\
& = \frac{128\pi^4}{\lambda_4(\lambda_3+\lambda_4)^2(\lambda_1^2-\lambda_{12}^2)} ln \left(\frac{\lambda_2+\lambda_{12}+\lambda_3+\lambda_4}{\lambda_1+\lambda_2+\lambda_3+\lambda_4} \times \frac{\lambda_1+\lambda_2}{\lambda_2+\lambda_{12}}\right),
\end{split}
\end{align}
which, in the limit $\lambda_{12}\rightarrow\lambda_1$, reduces to  equation(94) exactly, as expected. Taking $\lambda_{34}=0$ on both sides of  equation(93), a closed-form expression is obtained for the following integral :

\begin{align}  %Equation-97
\begin{split}
&K_{41}^g(\lambda_1, \lambda_2, \lambda_3, \lambda_4; \lambda_1, 0, 0) \\
& = \frac{128\pi^4}{\lambda_1(\lambda_1+\lambda_2)^2\lambda_4^2} ln \left(\frac{\lambda_1+\lambda_2+\lambda_3}{\lambda_1+\lambda_2+\lambda_3+\lambda_4} \times \frac{\lambda_3+\lambda_4}{\lambda_3}\right)\cdot
\end{split}
\end{align}
Substituting $\lambda_{12}=0$ on both sides of  equation(96), the following integral is evaluated :
\begin{align}  %Equation-98
\begin{split}
&K_{41}^g(\lambda_1, \lambda_2, \lambda_3, \lambda_4; 0, 0, \lambda_4) \\
& = \frac{128\pi^4}{\lambda_4\lambda_1^2(\lambda_3+\lambda_4)^2} ln \left(\frac{\lambda_2+\lambda_3+\lambda_4}{\lambda_1+\lambda_2+\lambda_3+\lambda_4} \times \frac{\lambda_1+\lambda_2}{\lambda_2}\right)\cdot
\end{split}
\end{align}
If $\lambda_1=\lambda_2=\lambda_3=\lambda_4=\delta$, then  equations(97) and (98) simplify to 
\begin{align}  %Equation-99
K_{41}^g(\delta, \delta, \delta, \delta; \delta, 0, 0) = K_{41}^g(\delta, \delta, \delta, \delta; 0, 0, \delta) =32 \pi^4~ ln (3/2)\delta^{-5} \cdot
\end{align}
The integral defined by
\begin{align}  %Equation-100
\begin{split}
&K_{41}(-1, -1, 0, -1 ; -1, -1, -1; \lambda_1, \lambda_2, \lambda_3, \lambda_4 ; \lambda_{12}, \lambda_{23}, \lambda_{34}) \\
&  = \int{d\overrightarrow{r_1}d\overrightarrow{r_2}d\overrightarrow{r_3}d\overrightarrow{r_4} (r_1r_2r_4r_{12}r_{23}r_{34})^{-1} }\\
 &\times exp (-\lambda_1r_1-\lambda_2r_2-\lambda_3r_3-\lambda_4r_4-\lambda_{12}r_{12}-\lambda_{23}r_{23}-\lambda_{34}r_{34}),
\end{split}
\end{align}
has been analytically evaluated, as a special case, by differentiating both sides of  equation(88) with respect to $(-\lambda_3)$. The property mentioned in  equation(28) is employed for the differentiation of the $dilog$ functions. Proceeding this way and doing some lengthy, but straightforward algebra, an expression is obtained which is shown to be exactly identical with the expression reported earlier in  equation(32) of paper I, wherein the integral was evaluated directly. Further, as pointed out in [59],  equation(32) of paper I is also reproduced, as a special case, by taking $(-\partial/\partial\lambda_3)$ of  equation(52) of [59], which gives an alternative expression for the integral $K_{41}^g$ , reported very recently.

By taking $(-\partial/\partial\lambda_2)(\partial^2/\partial\lambda_{23}^2)$  of the integral in  equation(100) and its closed - form expression in  equation(32) of paper I, and then substituting $\lambda_{12}=\lambda_{23}=\lambda_{34}=0$,  the following integral is evaluated as a special case :
\begin{align}  %Equation-101
\begin{split}
&K_{41}(-1, 0, 0, -1 ; -1, 1,-1 ; \lambda_1, \lambda_2, \lambda_3, \lambda_4 ; 0, 0, 0) \\
&= \frac{512\pi^4}{\lambda_1\lambda_4^2} \Bigg[ \frac{\lambda_3(\lambda_1+2\lambda_2+\lambda_3)}{\lambda_3^4\lambda_2(\lambda_1+\lambda_2)(\lambda_2+\lambda_3)(\lambda_1+\lambda_2+\lambda_3)} \\& \kern 7pc +\frac{\lambda_4(2\lambda_3+\lambda_4)(\lambda_1^2+3\lambda_1\lambda_2+3\lambda_2^2)}{\lambda_3^2(\lambda_3+\lambda_4)^2\lambda_2^3(\lambda_1+\lambda_2)^3} \\
&\kern 3pc -\frac{\lambda_3(\lambda_1+\lambda_3+2\lambda_2+2\lambda_4) + \lambda_4(\lambda_1+\lambda_4+2\lambda_2)}{(\lambda_3+\lambda_4)^4\lambda_2(\lambda_1+\lambda_2)(\lambda_2+\lambda_3+\lambda_4)(\lambda_1+\lambda_2+\lambda_3+\lambda_4) } \Bigg]\cdot
\end{split}
\end{align}

A different expression for the above integral, obtained by an alternative method, has been reported in  equation(51) of [52]. However, it has been verified that both the expressions yield the same number for the same set of values of $\lambda_1, \lambda_2, \lambda_3$ and $\lambda_4$  taken in both cases for calculation. Further, in the most special case, if $\lambda_1=\lambda_2=\lambda_3=\lambda_4=\delta$,   equation(101) also simplifies to  equation(53) of [52],indicating that the derivation leading to  equation(101) is correct.

An expression for another integral of interest, defined by 
\begin{align}  %Equation-102
\begin{split}
&K_{41}(-1, 0, 0, -1 ; -1, -1, -1; \lambda_1, \lambda_2, \lambda_3, \lambda_4 ; \lambda_{12}, \lambda_{23}, \lambda_{34}) \\
& = \int{d\overrightarrow{r_1}d\overrightarrow{r_2}d\overrightarrow{r_3}d\overrightarrow{r_4} (r_1~r_4~r_{12}~r_{23}~r_{34})^{-1} }\\
& \times exp (-\lambda_1r_1-\lambda_2r_2-\lambda_3r_3-\lambda_4r_4-\lambda_{12}r_{12}-\lambda_{23}r_{23}-\lambda_{34}r_{34}),
\end{split}
\end{align}
has been obtained, as a special case, by differentiating the integral in  equation(100) and its closed - form expression in  equation(32) of paper I, with respect to $(-\lambda_2)$,  and observed to be in conformity with  equation(11) of the paper of Bonham [63], which was pointed out earlier in paper I. Substituting $\lambda_{12}=\lambda_{23}=\lambda_{34}=0$ in the closed-form expression for the integral defined in  equation(102), the expressions in  equation(37), and, as a special case in  equation(38) of [52] are exactly reproduced.

All these observations point out that the derivation of the general expression leading to  equation(88) is correct.

\textbf{Expression for $ < r_{12}~r_{34}/r_{23}>$ }

As pointed out relating to  equation(78), it may also be required to obtain the expectation value $ < r_{12}~r_{34}/r_{23}>$ employing the expression for the generating integral $K_{41}^g$ as per  equation(88). This can be achieved by differentiating both sides of  equation(88) twice with respect to $\lambda_{12}$ and twice with respect to $\lambda_{34}$ and making use of  equation(28) for the differentiation of the $dilog$ functions. Thus the following $K_{41}$ integral is evaluated analytically :
\begin{align}  %Equation-103
\begin{split}
&K_{41}(-1, -1, -1, -1, ; 1, -1, 1 ; \lambda_1, \lambda_2, \lambda_3, \lambda_4 ; \lambda_{12}, \lambda_{23}, \lambda_{34}) \\
& = \frac{\partial^2}{\partial\lambda_{12}^2} \frac{\partial^2}{\partial\lambda_{34}^2}\bigg[D(\lambda_1,\lambda_{12},\lambda_4,\lambda_{34}) \times N(\lambda_1,\lambda_2,\lambda_3,\lambda_4 ; \lambda_{12},\lambda_{23}, \lambda_{34})\bigg],
\end{split}
\end{align}
where D and N are functions given by  equation(88) along with their expressions in  equations(86) and (89), respectively.

In the limit $\lambda_{12}=\lambda_{23}=\lambda_{34} \to 0$ ,  equation(103) is rewritten as 
\begin{align}  %Equation-104
\begin{split}
&K_{41}(-1, -1, -1, -1 ; 1, -1, 1 ; \lambda_1, \lambda_2, \lambda_3, \lambda_4 ; 0, 0, 0) \\
& = \lim_{\lambda_{12} \to 0},\lim_{\lambda_{34} \to 0}\frac{\partial^2}{\partial\lambda_{12}^2} \frac{\partial^2}{\partial\lambda_{34}^2} K_{41}^g(\lambda_1,\lambda_2,\lambda_3,\lambda_4 ; \lambda_{12}, 0, \lambda_{34}),
\end{split}
\end{align}
where the analytic expression for $K_{41}^g$ above is given by  equation(90). The differentiations, though lengthy but straightforward, are carried out and then the limits taken to obtain the following closed-form expression :
\begin{align}  %Equation-105
\begin{split}
&K_{41}(-1, -1, -1, -1 ; 1, -1, 1 ; \lambda_1, \lambda_2, \lambda_3, \lambda_4 ; 0, 0, 0) \\
& = 256\pi^4\left[4(\lambda_1\lambda_4)^{-4}F_1+2\lambda_4^{-4}\lambda_1^{-2}F_2+2\lambda_1^{-4}\lambda_4^{-2}F_3+(\lambda_1\lambda_4)^{-2}F_4\right],
\end{split}
\end{align}
where $F_1,\, F_2,\, F_3$ and $F_4$ are functions given by 
\begin{align*}
\begin{split}
&F_1=\frac{1}{\lambda_1+\lambda_2} ln \bigg(\frac{\lambda_1+\lambda_2+\lambda_3+\lambda_4}{\lambda_3+\lambda_4} \times \frac{\lambda_3}{\lambda_1+\lambda_2+\lambda_3}\bigg)\\
&+\frac{1}{\lambda_3+\lambda_4} ln \bigg(\frac{\lambda_1+\lambda_2+\lambda_3+\lambda_4}{\lambda_1+\lambda_2} \times \frac{\lambda_2}{\lambda_2+\lambda_3+\lambda_4}\bigg)\\
&+ \frac{1}{\lambda_2} ln \left(  \frac{\lambda_2+\lambda_3}{\lambda_3} \times \frac{\lambda_3+\lambda_4}{\lambda_2+\lambda_3+\lambda_4}\right)+\frac{1}{\lambda_3} ln \left(  \frac{\lambda_2+\lambda_3}{\lambda_2} \times \frac{\lambda_1+\lambda_2}{\lambda_1+\lambda_2+\lambda_3}\right), \\
%%%%%%%%%%%%%%%%
&F_2=\frac{1}{\lambda_3+\lambda_4} \bigg\{\frac{1}{(\lambda_2+\lambda_3+\lambda_4)^2} - \frac{1}{\lambda_2^2}\bigg\} + \frac{2}{\lambda_2^3} ln \left(\frac{\lambda_2+\lambda_3}{\lambda_3} \times \frac{\lambda_3+\lambda_4}{\lambda_2+\lambda_3+\lambda_4}\right) \\
&+\frac{2}{\lambda_2^2} \left(  \frac{1}{\lambda_2+\lambda_3+\lambda_4}-\frac{1}{\lambda_2+\lambda_3}\right) + \frac{1}{\lambda_2}\bigg\{ \frac{1}{(\lambda_2+\lambda_3+\lambda_4)^2}-\frac{1}{(\lambda_2+\lambda_3)^2}\bigg\}\\&+\frac{1}{\lambda_3} \bigg\{\frac{1}{\lambda_2^2}-\frac{1}{(\lambda_2+\lambda_3)^2} \bigg\},\\
%%%%%%%%%%%%%%%%%%%%%
&F_3=\frac{1}{\lambda_1+\lambda_2} \bigg\{ \frac{1}{(\lambda_1+\lambda_2+\lambda_3)^2} - \frac{1}{\lambda_3^2}\bigg\} + \frac{2}{\lambda_3^3} ln \left(\frac{\lambda_2+\lambda_3}{\lambda_2} \times \frac{\lambda_1+\lambda_2}{\lambda_1+\lambda_2+\lambda_3}\right) \\
&+\frac{2}{\lambda_3^2} \left(  \frac{1}{\lambda_1+\lambda_2+\lambda_3}-\frac{1}{\lambda_2+\lambda_3}\right) + \frac{1}{\lambda_3}\bigg\{ \frac{1}{(\lambda_1+\lambda_2+\lambda_3)^2}-\frac{1}{(\lambda_2+\lambda_3)^2}\bigg\}\\
&+\frac{1}{\lambda_2} \bigg\{\frac{1}{\lambda_3^2}-\frac{1}{(\lambda_2+\lambda_3)^2} \bigg\},\\
%%%%%%%%%%%%%%%%%%%%%%%%%%%%%%%%
&F_4=\frac{2}{\lambda_2^3} \bigg\{ \frac{1}{\lambda_3^2} - \frac{1}{(\lambda_2+\lambda_3)^2}\bigg\} + \frac{2}{\lambda_3^3} \bigg\{ \frac{1}{\lambda_2^2} - \frac{1}{(\lambda_2+\lambda_3)^2} \bigg\}-\frac{4}{\lambda_2^2(\lambda_2+\lambda_3)^3} \\
&- \frac{4}{\lambda_3^2(\lambda_2+\lambda_3)^3}-\frac{6}{\lambda_2(\lambda_2+\lambda_3)^4}-\frac{6}{\lambda_3(\lambda_2+\lambda_3)^4}\cdot
\end{split}
\end{align*}
In the most special case, if $\lambda_1=\lambda_2=\lambda_3=\lambda_4=\delta$, then  equation(105) simplifies to 
\begin{equation} % Equation 106
\begin{split}
&K_{41}(-1, -1, -1, -1; 1, -1, 1; \delta, \delta, \delta, \delta ; 0, 0, 0) \\
&= 256\pi^4\delta^{-9}[7/12~+~36~ln2~~-20~~ln3]\cdot
\end{split}
\end{equation}

\textbf{Expression for $<r_{12}~r_{23} / r_{34}>$ } 

By employing the generating integral $K_{41}^g$, as usual, an expression for $<r_{12} ~r_{23} / r_{34}>$ can be obtained in general. Thus, both sides of  equation(88) are differentiated twice with respect to $\lambda_{12} $ and twice with respect to $\lambda_{23} $ for evaluating the following integral analytically :
\begin{align} %Equation 107
\begin{split}
&K_{41}(-1, -1, -1, -1 ; 1, 1, -1 ; \lambda_1, \lambda_2, \lambda_3, \lambda_4 ; \lambda_{12}, \lambda_{23}, \lambda_{34})\\
&= \left(\frac{\partial^2}{\partial\lambda_{12}^2}\right)\left(\frac{\partial^2}{\partial\lambda_{23}^2}\right)\bigg[D(\lambda_1, \lambda_{12}, \lambda_4, \lambda_{34}) \times N(\lambda_1, \lambda_2, \lambda_3, \lambda_4 ; \lambda_{12}, \lambda_{23}, \lambda_{34})\bigg]\cdot
\end{split}
\end{align}
Expressions for the functions D and N are given by  equations(86) and (89), respectively. Differentiation of the $dilog$ functions in N can be performed by employing  equation(28).

To simplify  equation(107) further, the following observations are made, namely, (i) the function D does not depend on $\lambda_{23}$, (ii) as per  equation(85), the function N is a sum of four different $I_2(-2, -2, -1 ; \alpha, \beta, \gamma)$ integrals and the relations of various $\alpha, \beta, \gamma$ parameters with $\lambda$ parameters are given along with  equation(87), (iii) the parameter $\gamma$  for all the four $I_{2}$ integrals is equal to $\lambda_{23}$,  and (iv) employing  equation(87), it is easy to establish the relation
\begin{equation} %Equiation - 108
\left(\frac{\partial^2}{\partial\gamma^2}\right)I_2(-2, -2, -1 ; \alpha, \beta, \gamma)=I_2(-2, -2, 1 ; \alpha, \beta, \gamma)\cdot
\end{equation}
Accordingly,  equation(107) simplifies to give
\begin{align} %Equation 109
\begin{split}
& K_{41}(-1, -1, -1, -1 ; 1, 1, -1 ; \lambda_1, \lambda_2, \lambda_3, \lambda_4 ; \lambda_{12}, \lambda_{23}, \lambda_{34})= \left(\frac{\partial^2}{\partial\lambda_{12}^2}\right)\\
&\times \bigg[D(\lambda_1, \lambda_{12}, \lambda_4, \lambda_{34})\bigg\{I_2(-2, -2, 1; \alpha_1, \beta_1, \lambda_{23})-I_2(-2, -2, 1; \alpha_2, \beta_2, \lambda_{23}) \\
&-I_2(-2, -2, 1; \alpha_3, \beta_3, \lambda_{23}) +I_2(-2, -2, 1; \alpha_4, \beta_4, \lambda_{23})\bigg\}\bigg]\cdot
\end{split}
\end{align}

Closed - form expression for $I_2(-2, -2, -1; \alpha, \beta, \gamma)$ has been given in  equation(25), and hence, through  equation(108), all the four $I_2(-2, -2, 1 ; \alpha, \beta, \gamma)$  integrals in  equation(109) can be evaluated analytically. Then the differentiation with respect to $\lambda_{12}$ is performed twice keeping in mind that the function D and only the parameters $\alpha_1$ and $\alpha_2$ involve $\lambda_{12}$.

\textbf{Limiting expression if $\lambda_{12}=\lambda_{23}=\lambda_{34}=0$}

First an expression for the integral $I_2(-2, -2, 1; \alpha, \beta, \gamma)$ in the limit $\gamma \to 0$ is obtained as given below.

Substituting  equation(25) in  equation(108), it becomes
\begin{align}%%Equation 110
\begin{split}
&I_2(-2, -2, 1; \alpha, \beta, \gamma) = \left( \frac{\partial^2}{\partial\gamma^2} \right)\bigg[8 \pi^2 \gamma^{-1} Q(\alpha, \beta, \gamma)\bigg],
\end{split}
\end{align}
with the expression for the function $Q(\alpha, \beta, \gamma)$ given by  equation(26). Performing the differentiation on the right hand side,  equation(110) reduces to
\begin{align}%%Equation 111
\begin{split}
I_2(-2, -2, 1; \alpha, \beta, \gamma) 
= \frac{8\pi^2}{\gamma^3} \times \bigg[2Q-2\gamma\left(\frac{\partial Q}{\partial\gamma}\right)+\gamma^2 \left(\frac{\partial^2 Q}{\partial\gamma^2}\right)\bigg]\cdot
\end{split}
\end{align}
It had been shown [ ] that $Q\to 0$ as $\gamma \to 0$. However, $\left(\frac{\partial Q}{\partial \gamma}\right)$ and $\left(\frac{\partial^2 Q}{\partial \gamma^2}\right)$ both can be observed to be finite as $\gamma \to 0$.  Thus in the limit $\gamma \to 0$ , the right hand side expression in  equation(111) assumes $0/0$ form. Hence L' Hospital's rule for $0/0$ is employed to obtain
\begin{align}%%Equation 112
\begin{split}
&I_2(-2, -2, 1; \alpha, \beta, 0) = \left(\frac{8\pi^2}{3}\right)\left(\frac{\partial^3 Q}{\partial\gamma^3}\right)|_{\gamma=0}\\
&=\frac{16\pi^2}{3}\bigg[\frac{\alpha+\beta}{\alpha^2\beta^2}+\frac{2}{\beta^3}ln\left(\frac{\alpha+\beta}{\alpha}\right)+\frac{2}{\alpha^3}ln \left(\frac{\alpha+\beta}{\beta}\right)\bigg]\cdot
\end{split}
\end{align}
Accordingly, if $\lambda_{23} = 0$,  equation(109) is restated as 
\begin{align}%%Equation 113
\begin{split}
&K_{41}(-1, -1, -1, -1; 1, 1, -1; \lambda_1, \lambda_2, \lambda_3, \lambda_4 ; \lambda_{12}, 0, \lambda_{34}) =\left(\frac{\partial^2}{\partial\lambda_{12}^2} \right)\\
&\times \bigg[ D(\lambda_1, \lambda_{12}, \lambda_4, \lambda_{34})\bigg\{I_2(-2, -2, 1; \alpha_1, \beta_1, 0) - I_2(-2, -2, 1; \alpha_2, \beta_2, 0) \\
& \kern 7pc - I_2(-2, -2, 1; \alpha_3, \beta_3, 0) + I_2 (-2, -2, 1; \alpha_4, \beta_4, 0)\bigg\}\bigg]\cdot
\end{split}
\end{align}
Expressing the function D, and $\alpha$'s and $\beta$'s, in terms of $\lambda$'s as per  equations(86) and (87), respectively, and inserting  equation(112) in  equation(113), the differentiation with respect to $\lambda_{12}$ is carried out twice. Then, substituting $\lambda_{12} = \lambda_{34}=0$ in the resulting expression, the integral in  equation(113) becomes 
\begin{align}%%Equation 114
K_{41}(-1, -1, -1, -1; 1, 1, -1; \lambda_1, \lambda_2, \lambda_3, \lambda_4; 0, 0, 0) 
=\frac{512}{3}\quad \frac{\pi^4}{\lambda_1^2\lambda_4^2} \sum_{i=1}^8 X_i,
\end{align}
where
\begin{align*}
\begin{split}
&X_1=\frac{1}{\lambda_1^2}\bigg\{\frac{\lambda_2+\lambda_3}{\lambda_2^2\lambda_3^2}+\frac{2}{\lambda_3^3}ln\left(\frac{\lambda_2+\lambda_3}{\lambda_2}\right)+\frac{2}{\lambda_2^3}ln\left(\frac{\lambda_2+\lambda_3}{\lambda_3}\right)\bigg\}, \\
&X_2=-\frac{1}{\lambda_1^2}\bigg\{\frac{\lambda_2+\lambda_3+\lambda_4}{\lambda_2^2(\lambda_3+\lambda_4)^2}+\frac{2}{(\lambda_3+\lambda_4)^3}ln\left(\frac{\lambda_2+\lambda_3+\lambda_4}{\lambda_2}\right)+\frac{2}{\lambda_2^3}ln\left(\frac{\lambda_2+\lambda_3+\lambda_4}{\lambda_3+\lambda_4}\right)\bigg\}, \\
&X_3=-\frac{1}{\lambda_1^2}\bigg\{\frac{\lambda_1+\lambda_2+\lambda_3}{\lambda_3^2(\lambda_1+\lambda_2)^2}+\frac{2}{\lambda_3^3}ln\left(\frac{\lambda_1+\lambda_2 +\lambda_3}{\lambda_1+\lambda_2}\right)  
+\frac{2}{(\lambda_1+\lambda_2)^3}ln\left(\frac{\lambda_1+\lambda_2+\lambda_3}{\lambda_3}\right)\bigg\}, \\
&X_4=\frac{1}{\lambda_1^2}\bigg\{\frac{\lambda_1+\lambda_2+\lambda_3+\lambda_4}{(\lambda_1+\lambda_2)^2(\lambda_3+\lambda_4)^2}  
+\frac{2}{(\lambda_3+\lambda_4)^3}ln\left(\frac{\lambda_1+\lambda_2+\lambda_3+\lambda_4}{\lambda_1+\lambda_2}\right),  \\
&+\frac{2}{(\lambda_1+\lambda_2)^3}ln\left(\frac{\lambda_1+\lambda_2+\lambda_3+\lambda_4}{\lambda_3+\lambda_4}\right)\bigg\}, \\
&X_5=-\frac{2}{\lambda_3^2\lambda_2^3}+\frac{3(\lambda_2+\lambda_3)}{\lambda_2^4\lambda_3^2}+\frac{1}{\lambda_3^3\lambda_2^2}  
-\frac{1}{\lambda_3^3(\lambda_2+\lambda_3)^2}, \\
&X_6=\frac{12}{\lambda_2^5} ln\left(\frac{\lambda_2+\lambda_3}{\lambda_3}\right)-\frac{6}{\lambda_2^4(\lambda_2+\lambda_3)}-\frac{1}{\lambda_2^3(\lambda_2+\lambda_3)^2}, \\
%%%%%%%%%%%
&X_7=\frac{2}{\lambda_2^3(\lambda_3+\lambda_4)^2}-\frac{3(\lambda_2+\lambda_3+\lambda_4)}{\lambda_2^4(\lambda_3+\lambda_4)^2} 
-\frac{1}{(\lambda_3+\lambda_4)^3} \bigg\{\frac{1}{\lambda_2^2}-\frac{1}{(\lambda_2+\lambda_3+\lambda_4)^2}\bigg\}, \\
%%%%%%%%%%%%
&X_8=-\frac{12}{\lambda_2^5}ln\left(\frac{\lambda_2+\lambda_3+\lambda_4)}{\lambda_3+\lambda_4} \right)+\frac{6}{\lambda_2^4(\lambda_2+\lambda_3+\lambda_4)}+\frac{1}{\lambda_2^3(\lambda_2+\lambda_3+\lambda_4)^2} \cdot 
\end{split}
\end{align*}
If $\lambda_1=\lambda_2=\lambda_3=\lambda_4=\delta$, then  equation(114) simplifies to 
\begin{equation} %Equation No.115
\begin{split}
&K_{41}(-1, -1, -1, -1 ; 1, 1, -1 ; \delta, \delta, \delta, \delta ; 0, 0, 0) \\
&=(256/3)\pi^4\delta^{-9}(5+65~ln2 - 33~ln3)\cdot
\end{split}
\end{equation}
Differentiating both sides of  equation(114) with respect to $(-\lambda_3)$, a closed-form expression for the integral
\begin{align}%%Equation 116
\begin{split}
&K_{41}(-1, -1, 0, -1; 1, 1, -1; \lambda_1, \lambda_2, \lambda_3, \lambda_4; 0, 0, 0)= \int {d\overrightarrow{r_1}d\overrightarrow{r_2}d\overrightarrow{r_3}d\overrightarrow{r_4}}~ \\
& \times (r_1r_2r_4)^{-1}(r_{12}r_{23}/r_{34}) \times exp(-\lambda_1r_1-\lambda_2r_2-\lambda_3r_3-\lambda_4r_4),
\end{split}
\end{align}
is easily obtained. Substituting $\lambda_1=\lambda_2=\lambda_3=\lambda_4=\delta$  in this expression, in the most special case, the following equation is established :
\begin{align}%%Equation 117
\begin{split}
&K_{41}(-1, -1, 0, -1; 1, 1, -1; \delta, \delta, \delta, \delta; 0, 0, 0) \\
&= \left(\frac{32\pi^4}{9}\right)\delta^{-10}[594~ln2 - 306~ln3 +505] \cdot
\end{split}
\end{align}
Further, by taking $\left(\frac{\partial}{\partial\lambda_4}\right)\left(\frac{\partial}{\partial\lambda_3}\right)$ of  equation(114), a closed-form expression for the following integral
\begin{equation}%%Equation 118
\begin{split}
&K_{41}(-1, -1, 0, 0 ; 1, 1, -1; \lambda_1, \lambda_2, \lambda_3, \lambda_4; 0, 0, 0) = \int {d\overrightarrow{r_1}~d\overrightarrow{r_2}~d\overrightarrow{r_3}~d\overrightarrow{r_4}} \\ 
&\kern 5pc \times (r_1r_2)^{-1}(r_{12}r_{23}/r_{34})~exp(-\lambda_1r_1-\lambda_2r_2-\lambda_3r_3-\lambda_4r_4)
\end{split} 
\end{equation}
is derived. If, in the most special case, $\lambda_1=\lambda_2=\lambda_3=\lambda_4=\delta$     substituted in the expression for the above integral, it yields the following equation :
\begin{align} %Equation 119
\begin{split}
&K_{41}(-1, -1, 0, 0 ; 1, 1, -1; \delta, \delta, \delta, \delta ; 0, 0, 0)\\&=\left(\frac{16\pi^4}{27}\right)\delta^{-11} \left[7344~ln2-3888~ln3+5167\right],
\end{split}
\end{align}
which is exactly identical with  equation(24) of the very recent report of King [59]. This observation clearly leads to the conclusion that the derivations leading to  equations(114), (115), (117) and (119) are correct.

\textbf{(b) Analytic evaluation of $K_{42}^g$ }

The generating integral $K_{42}^g$ is defined, as usual, by the relation
\begin{equation}% Equation 120
\begin{split}
&K_{42}^g(\lambda_{1}, \lambda_{2}, \lambda_3, \lambda_4 ; \lambda_{23}, \lambda_{34}, \lambda_{41}) = \int{d\overrightarrow{r_1}~d\overrightarrow{r_2}~d\overrightarrow{r_3}~d\overrightarrow{r_4}~(r_1~r_2~r_3~r_4~r_{23}~r_{34}~r_{41})^{-1}}\\&~~ \times~exp(-\lambda_1r_1-\lambda_2r_2-\lambda_3r_3-\lambda_4r_4-\lambda_{23}r_{23}-\lambda_{34}r_{34}-\lambda_{41}r_{41})\cdot
\end{split}
\end{equation}
Comparing the integrand in the above integral with that given for $K_{41}^g$ in  equation(83), it is observed that if a change $(2\leftrightarrows4)$ is performed in the integrand in  equation(83), the integrand on the right hand side of  equation(120) is obtained, since $\lambda_{ij}=\lambda_{ji}$ and $r_{ij}=r_{ji}$. Hence, by observation, the following equation is establised, starting from  equations(88) and (89) :
\begin{equation}%Equation 121
\begin{split}
&K_{42}^g(\lambda_1, \lambda_2, \lambda_3, \lambda_4; \lambda_{23},\lambda_{34},\lambda_{41})=\frac{16\pi^2}{(\lambda_2^2-\lambda_{23}^2)(\lambda_1^2-\lambda_{41}^2)}~\times~\frac{8\pi^2}{\lambda_{34}}\\
& \times \bigg[ln\left(\frac{\lambda_{41}+\lambda_4+\lambda_{34}}{\lambda_1+\lambda_4+\lambda_{34}}\right)~\times~ln\left(\frac{\lambda_3+\lambda_2+\lambda_{34}}{\lambda_3+\lambda_{23}+\lambda_{34}}\right) \\
&+dilog \left(\frac{\lambda_4+\lambda_{41}+\lambda_3+\lambda_{23}}{\lambda_4+\lambda_{41}+\lambda_{34}}\right) + dilog \left(\frac{\lambda_4+\lambda_{41}+\lambda_3+\lambda_{23}}{\lambda_3+\lambda_{23}+\lambda_{34}}\right)\\
&-dilog \left(\frac{\lambda_4+\lambda_{41}+\lambda_3+\lambda_2}{\lambda_4+\lambda_{41}+\lambda_{34}}\right) - dilog \left(\frac{\lambda_4+\lambda_{41}+\lambda_3+\lambda_2}{\lambda_3+\lambda_2+\lambda_{34}}\right)\\
&-dilog \left(\frac{\lambda_1+\lambda_4+\lambda_3+\lambda_{23}}{\lambda_1+\lambda_4+\lambda_{34}}\right) - dilog \left(\frac{\lambda_1+\lambda_4+\lambda_3+\lambda_{23}}{\lambda_3+\lambda_{23}+\lambda_{34}}\right)\\
&+dilog \left(\frac{\lambda_1+\lambda_4+\lambda_3+\lambda_2}{\lambda_1+\lambda_4+\lambda_{34}}\right) + dilog \left(\frac{\lambda_1+\lambda_4+\lambda_3+\lambda_2}{\lambda_3+\lambda_2+\lambda_{34}}\right)\bigg]\cdot
\end{split}
\end{equation}

It is worth pointing out here that the integral in  equation(120) has also been evaluated directly adopting the same procedure as followed for the evaluation of the integral $K_{41}^g$ defined in  equation(83), and subsequently,  equation(121) is established as outlined below.

The integral in  equation(120) can be rewritten  as
\begin{equation}%Equation 122
\begin{split}
&K_{42}^g(\lambda_1, \lambda_2, \lambda_3, \lambda_4; \lambda_{23},\lambda_{34},\lambda_{41})=\int{d\overrightarrow{r_3}~d\overrightarrow{r_4}(r_3r_4r_{34})^{-1}}\\
&\times exp(-\lambda_3r_3-\lambda_4r_4-\lambda_{34}r_{34})~J(\lambda_2,\lambda_{23},r_3)~J(\lambda_1,\lambda_{41},r_4),
\end{split}
\end{equation}

where the $J$'s are given by  equations(1) and (3). Substituting the closed-form expressions for the $J$'s in  equation(122), it simplifies to 
\begin{equation}%Equation 123
\begin{split}
&K_{42}^g(\lambda_1, \lambda_2, \lambda_3, \lambda_4; \lambda_{23}, \lambda_{34}, \lambda_{41})=\frac{16\pi^2}{(\lambda_2^2-\lambda_{23}^2)(\lambda_1^2-\lambda_{41}^2)}\\
&\times\bigg[I_2(-2,-2,-1; \lambda_3+\lambda_{23}, \lambda_4+\lambda_{41}, \lambda_{34}) 
-I_2(-2,-2,-1; \lambda_3+\lambda_{23}, \lambda_1+\lambda_4, \lambda_{34})\\
&-I_2(-2,-2,-1; \lambda_2+\lambda_3, \lambda_4+\lambda_{41}, \lambda_{34}) +I_2(-2,-2,-1; \lambda_2+\lambda_3, \lambda_1+\lambda_4, \lambda_{34})\bigg],
\end{split}
\end{equation}
where, in general, the closed-form expression for the $I_2$ integral is given by  equation(25). Substituting these expressions for various $I_2$ integrals in  equation(123), and doing some simplifications, the expression in  equation(121) is exactly reproduced.

Further, replacing the various $\alpha,~\beta$ and $\gamma$ parameters in the right hand side expression in  equation(85) by the corresponding $\lambda$ parameters given just after  equation(87), and then making a change $(2\leftrightarrows4)$ in the resulting expression, the right hand side expression in  equation(123) is exactly reproduced, as expected.

\textbf{Limiting expressions for $K_{42}^g$}

In the limit $\lambda_{34} \to 0$, employing  equation(29), the following integral is evaluated, as a special case, from  equation(123) :
\begin{equation} %Equation No.124
\begin{split}
&K_{42}^g(\lambda_1, \lambda_2, \lambda_3, \lambda_4; \lambda_{23}, 0, \lambda_{41}) =\frac{256\pi^4}{(\lambda_2^2-\lambda_{23}^2)(\lambda_1^2-\lambda_{41}^2)} \\
&\times\bigg[ \frac{1}{\lambda_1+\lambda_4}ln\left(\frac{\lambda_2+\lambda_3+\lambda_1+\lambda_4}{\lambda_2+\lambda_3}~\times~\frac{\lambda_3+\lambda_{23}}{\lambda_3+\lambda_{23}+\lambda_1+\lambda_4}\right)\\
&+\frac{1}{\lambda_2+\lambda_3}ln\left(\frac{\lambda_2+\lambda_3+\lambda_1+\lambda_4}{\lambda_1+\lambda_4}~\times~\frac{\lambda_4+\lambda_{41}}{\lambda_2+\lambda_3+\lambda_4+\lambda_{41}}\right)\\
&+\frac{1}{\lambda_4+\lambda_{41}}ln\left(\frac{\lambda_3+\lambda_{23}+\lambda_4+\lambda_{41}}{\lambda_3+\lambda_{23}}~\times~\frac{\lambda_2+\lambda_3}{\lambda_2+\lambda_3+\lambda_4+\lambda_{41}}\right)\\
&+\frac{1}{\lambda_3+\lambda_{23}}ln\left(\frac{\lambda_3+\lambda_{23}+\lambda_4+\lambda_{41}}{\lambda_4+\lambda_{41}}~\times~\frac{\lambda_1+\lambda_4}{\lambda_3+\lambda_{23}+\lambda_1+\lambda_4}\right)\bigg]\cdot
\end{split}
\end{equation}
Looking at the right hand side expressions in  equations(90) and (124), it is observed that the latter expression can  be obtained from the former by performing the change $(2\leftrightarrows4)$ in the former, as expected. Also expressions for the integrals $K_{42}^g(\lambda_1, \lambda_2, \lambda_3, \lambda_4; \lambda_{23}, 0, 0)$ and $K_{42}^g(\lambda_1, \lambda_2, \lambda_3, \lambda_4; 0, 0, \lambda_{41})$ can be easily obtained from  equation(124) by substituting $\lambda_{41}=0$ and $\lambda_{23}=0$, respectively, on both sides. If both $\lambda_{41}=\lambda_{23}=0$, then  equation(124) simplifies to give
\begin{equation} %Equation No.125
\begin{split}
&K_{42}^g(\lambda_1, \lambda_2, \lambda_3, \lambda_4; 0, 0, 0) \\
&=\frac{256\pi^4}{(\lambda_2^2\lambda_1^2)}\bigg[ \frac{1}{\lambda_1+\lambda_4}ln\left(\frac{\lambda_2+\lambda_3+\lambda_1+\lambda_4}{\lambda_2+\lambda_3}~\times~\frac{\lambda_3}{\lambda_3+\lambda_4+\lambda_1}\right)\\
&+\frac{1}{\lambda_2+\lambda_3}ln\left(\frac{\lambda_2+\lambda_3+\lambda_1+\lambda_4}{\lambda_1+\lambda_4}~\times~\frac{\lambda_4}{\lambda_2+\lambda_3+\lambda_4}\right)\\
&+\frac{1}{\lambda_4}ln\left(\frac{\lambda_3+\lambda_4}{\lambda_3}~\times~\frac{\lambda_2+\lambda_3}{\lambda_2+\lambda_3+\lambda_4}\right)+\frac{1}{\lambda_3}ln\left(\frac{\lambda_3+\lambda_4}{\lambda_4}~\times~\frac{\lambda_1+\lambda_4}{\lambda_3+\lambda_1+\lambda_4}\right)\bigg]\cdot
\end{split}
\end{equation}
Comparing the right hand side expressions in  equations(91) and (125), it is clearly observed that the expression in  equation(125) can also be obtained by making the change$(2\leftrightarrows4)$ in the expression in  equation(91), as expected. In the most special case, if $\lambda_1=\lambda_2=\lambda_3=\lambda_4=\delta$, then  equation(125) reduces to
\begin{equation} %Equation No.126
K_{42}^g(\delta, \delta, \delta, \delta; 0, 0, 0)=256\pi^4(5~ln2-3~ln3)\delta^{-5}\cdot
\end{equation}
Starting with  equation(124), several limiting expressions can be obtained, by employing L'Hospital's rule for $0/0$, as was done in case of $K_{41}^g$. Thus, in the limit $\lambda_{41} \to \lambda_1$,  equation(124) simplifies to 
\begin{equation} %Equation No.127
\begin{split}
&K_{42}^g(\lambda_1, \lambda_2, \lambda_3, \lambda_4; \lambda_{23}, 0,\lambda_1)\\
&=\frac{128\pi^4}{\lambda_1(\lambda_2^2-\lambda_{23}^2)(\lambda_4+\lambda_1)^2}~ln\left(\frac{\lambda_3+\lambda_{23}+\lambda_4+\lambda_1}{\lambda_2+\lambda_3+\lambda_4+\lambda_1}\times \frac{\lambda_2+\lambda_3}{\lambda_3+\lambda_{23}}\right)\cdot
\end{split}
\end{equation}
If, further, $\lambda_{23} \to \lambda_2$,  equation(127) becomes
\begin{equation} %Equation No.128
\begin{split}
&K_{42}^g(\lambda_1, \lambda_2, \lambda_3, \lambda_4; \lambda_2, 0,\lambda_1)\\
&=\frac{64\pi^4}{\lambda_1\lambda_2(\lambda_1+\lambda_4)(\lambda_2+\lambda_3)}~\times \frac{1}{\lambda_1+\lambda_2+\lambda_3+\lambda_4}\cdot
\end{split}
\end{equation}
If $\lambda_{23} \to \lambda_2$ first, then  equation(124) gives
\begin{equation} %Equation No.129
\begin{split}
&K_{42}^g(\lambda_1, \lambda_2, \lambda_3, \lambda_4; \lambda_2, 0,\lambda_{41})\\
&=\frac{128\pi^4}{\lambda_2(\lambda_2+\lambda_3)^2(\lambda_1^2-\lambda_{41}^2)}~ln\left(\frac{\lambda_2+\lambda_3+\lambda_4+\lambda_{41}}{\lambda_1+\lambda_2+\lambda_3+\lambda_4}\times \frac{\lambda_1+\lambda_4}{\lambda_4+\lambda_{41}}\right)\cdot
\end{split}
\end{equation}
If, further, $\lambda_{41} \to \lambda_1$, then  equation(129) reduces to  equation(128) exactly, as expected. Substituting $\lambda_{23}=0$ in  equation(127), the following integral is evaluated :
\begin{equation} %Equation No.130
\begin{split}
&K_{42}^g(\lambda_1, \lambda_2, \lambda_3, \lambda_4; 0, 0,\lambda_1)\\
&=\frac{128\pi^4}{\lambda_1\lambda_2^2(\lambda_4+\lambda_1)^2}~ln\left(\frac{\lambda_3+\lambda_4+\lambda_1}{\lambda_2+\lambda_3+\lambda_4+\lambda_1}\times \frac{\lambda_2+\lambda_3}{\lambda_3}\right)\cdot
\end{split}
\end{equation}
If  $\lambda_{41}=0$,  equation(129) simplifies to
\begin{equation} %Equation No.131
\begin{split}
&K_{42}^g(\lambda_1, \lambda_2, \lambda_3, \lambda_4; \lambda_2, 0,0)\\
&=\frac{128\pi^4}{\lambda_2(\lambda_2+\lambda_3)^2\lambda_1^2}~ln\left(\frac{\lambda_2+\lambda_3+\lambda_4}{\lambda_1+\lambda_2+\lambda_3+\lambda_4}\times \frac{\lambda_1+\lambda_4}{\lambda_4}\right)\cdot
\end{split}
\end{equation}
If $\lambda_1=\lambda_2=\lambda_3=\lambda_4=\delta$, then  equation(128) becomes
\begin{equation} %Equation No.132
\begin{split}
K_{42}^g(\delta, \delta, \delta, \delta ; \delta, 0, \delta) = 4\pi^4/\delta^5,
\end{split}
\end{equation}
and  equations(130) and (131) simplify to give
\begin{equation} %Equation No.133
K_{42}^g(\delta, \delta, \delta, \delta ; 0, 0, \delta) = K_{42}^g(\delta, \delta, \delta, \delta ; \delta, 0, 0) = 32\pi^4~ln(3/2)\delta^{-5}
\end{equation}
analogous with  equation(99).

\textbf{(c) Analytic evaluation of $K_{43}^g$}

The generating integral $K_{43}^g$ is defined by the equation
\begin{equation} %Equation No.134
\begin{split}
&K_{43}^g(\lambda_1, \lambda_2, \lambda_3, \lambda_4; \lambda_{34}, \lambda_{41},\lambda_{12})\\
&=\int{d\overrightarrow{r_1}d\overrightarrow{r_2}d\overrightarrow{r_3}d\overrightarrow{r_4}} (r_1r_2r_3r_4r_{34}r_{41}r_{12})^{-1} \\
&\times~exp(-\lambda_1r_1-\lambda_2r_2-\lambda_3r_3-\lambda_4r_4-\lambda_{34}r_{34}-\lambda_{41}r_{41}-\lambda_{12}r_{12}),
\end{split}
\end{equation}
which can be recast as
\begin{equation} %Equation No.135
\begin{split}
&K_{43}^g=\int{d\overrightarrow{r_1}d\overrightarrow{r_4}} (r_1r_4r_{41})^{-1} ~exp(-\lambda_1r_1-\lambda_4r_4-\lambda_{41}r_{41})\\
& \kern 3pc \times~J(\lambda_2, \lambda_{12},r_1)~J(\lambda_3, \lambda_{34},r_4),
\end{split}
\end{equation}
where the $J$'s are given by   equations(1) and (3). Substituting the expressions for the $J$'s from  equation(3) in the above equation, the following expression is obtained for $K_{43}^g$ :
\begin{equation} %Equation No.136
\begin{split}
&K_{43}^g(\lambda_1, \lambda_2, \lambda_3, \lambda_4; \lambda_{34}, \lambda_{41},\lambda_{12})=\frac{16\pi^2}{(\lambda_2^2-\lambda_{12}^2)(\lambda_3^2-\lambda_{34}^2)}\\
&\times\Big[I_2(-2,-2,-1;\lambda_1+\lambda_{12}, \lambda_4+\lambda_{34}, \lambda_{41})-I_2(-2,-2,-1;\lambda_1+\lambda_{12}, \lambda_3+\lambda_4, \lambda_{41})\\
&-I_2(-2,-2,-1;\lambda_1+\lambda_2, \lambda_4+\lambda_{34}, \lambda_{41})+I_2(-2,-2,-1;\lambda_1+\lambda_2, \lambda_3+\lambda_4, \lambda_{41})\Big]\cdot
\end{split}
\end{equation}
The general expression given for $I_2(-2,-2,-1; \alpha, \beta, \gamma)$ in  equation(25) can be employed in  equation(136) to obtain the required closed-form expression for $K_{43}^g$. Various limiting expressions for $K_{43}^g$ can be obtained as was done for $K_{41}^g$.

\textbf{(d) Analytic evaluation of $K_{44}^g$}

The generating integral $K_{44}^g$ is defined by the equation
\begin{equation} %Equation No.137
\begin{split}
&K_{44}^g(\lambda_1, \lambda_2, \lambda_3, \lambda_4; \lambda_{41}, \lambda_{12},\lambda_{23})=\int{d\overrightarrow{r_1}d\overrightarrow{r_2}d\overrightarrow{r_3}d\overrightarrow{r_4}} (r_1r_2r_3r_4r_{41}r_{12}r_{23})^{-1} \\
&\times~exp(-\lambda_1r_1-\lambda_2r_2-\lambda_3r_3-\lambda_4r_4-\lambda_{41}r_{41}-\lambda_{12}r_{12}-\lambda_{23}r_{23}),
\end{split}
\end{equation}
and is rewritten as 
\begin{equation} %Equation No.138
\begin{split}
&K_{44}^g=\int{d\overrightarrow{r_1}d\overrightarrow{r_2}} (r_1r_2r_{12})^{-1} ~exp(-\lambda_1r_1-\lambda_2r_2-\lambda_{12}r_{12})\\
&\times~J(\lambda_3, \lambda_{23},r_2)~J(\lambda_4, \lambda_{41},r_1)\cdot
\end{split}
\end{equation}
Inserting  equation(3) into  equation(138), the following expression for $K_{44}^g$ is obtained :
\begin{equation} %Equation No.139
\begin{split}
&K_{44}^g(\lambda_1, \lambda_2, \lambda_3, \lambda_4; \lambda_{41}, \lambda_{12},\lambda_{23})=\frac{16\pi^2}{(\lambda_3^2-\lambda_{23}^2)(\lambda_4^2-\lambda_{41}^2)}\\
&\times\Big[I_2(-2,-2,-1;\lambda_1+\lambda_{41}, \lambda_2+\lambda_{23}, \lambda_{12})-I_2(-2,-2,-1;\lambda_1+\lambda_{41}, \lambda_2+\lambda_3, \lambda_{12})\\
&-I_2(-2,-2,-1;\lambda_1+\lambda_4, \lambda_2+\lambda_{23}, \lambda_{12})+I_2(-2,-2,-1;\lambda_1+\lambda_4, \lambda_2+\lambda_3, \lambda_{12})\Big]\cdot
\end{split}
\end{equation}
Replacing the $I_2$ integrals in  equation(139) by their corresponding expressions as per  equation(25), closed-form expression for the  generating integral $K_{44}^g$ is easily obtained. Various limiting expressions for $K_{44}^g$ can be derived by following the same procedure as adopted for $K_{41}^g$.\\
\hfill\break
\textbf{5.2(II) Integrals belonging to category (II)}\\
\hfill\break
There are four general integrals belonging to category (II) of open square diagrams, corresponding to four different exponential correlation factors as mentioned earlier. These are denoted as (a)$K_{45}$, (b) $K_{46}$, (c)$K_{47}$ and (d) $K_{48}$, and the corresponding generating integrals as (a)$K_{45}^g$, (b) $K_{46}^g$, (c)$K_{47}^g$ and (d) $K_{48}^g$. The definition along with the method of analytic evaluation of each of these generating integrals is given below.

\textbf{(a) Analytic evaluation of $K_{45}^g$}

The generating integral $K_{45}^g$ is defined by the relation
\begin{equation} %Equation No.140
\begin{split}
&K_{45}^g(\lambda_1, \lambda_2, \lambda_3, \lambda_4 ; \lambda_{12}, \lambda_{13}, \lambda_{34}) = \int{d\overrightarrow{r_1}d\overrightarrow{r_2}d\overrightarrow{r_3}d\overrightarrow{r_4} (r_1r_2r_3r_4r_{12}r_{13} r_{34})^{-1}} \\
&\kern 3pc \times exp(-\lambda_1r_1 -\lambda_2r_2 -\lambda_3r_3 -\lambda_4r_4 -\lambda_{12}r_{12} -\lambda_{13}r_{13}- \lambda_{34}r_{34}),
\end{split}
\end{equation}
which can be recast as 
\begin{equation} %Equation No.141
\begin{split}
&K_{45}^g = \int{d\overrightarrow{r_1}d\overrightarrow{r_3}(r_1r_3r_{13})^{-1}} exp(-\lambda_1r_1 -\lambda_3r_3 -\lambda_{13}r_{13})\\
&\kern 7pc \times J(\lambda_2, \lambda_{12},r_1) J(\lambda_4, \lambda_{34},r_3),
\end{split}
\end{equation}
where expressions for the J's are given by  equation(3). Inserting  equation(3) in  equation(141),  the following equation is established :
\begin{equation} %Equation No.142
\begin{split}
&K_{45}^g(\lambda_1, \lambda_2, \lambda_3, \lambda_4; \lambda_{12}, \lambda_{13},\lambda_{34})=\frac{16\pi^2}{(\lambda_2^2-\lambda_{12}^2)(\lambda_4^2-\lambda_{34}^2)}\\
&\times\Big[I_2(-2,-2,-1;\lambda_1+\lambda_{12}, \lambda_3+\lambda_{34}, \lambda_{13})-I_2(-2,-2,-1;\lambda_1+\lambda_{12}, \lambda_3+\lambda_4, \lambda_{13})\\
&-I_2(-2,-2,-1;\lambda_1+\lambda_2, \lambda_3+\lambda_{34}, \lambda_{13})+I_2(-2,-2,-1;\lambda_1+\lambda_2, \lambda_3+\lambda_4, \lambda_{13})\Big],
\end{split}
\end{equation}
with expressions for the $I_2$ integrals given by  equation(25).

It is easy to observe that if a change $(1\leftrightarrows2)$ is made in the integrand in  equation(83) corresponding to $K_{41}^g$, the integrand for $K_{45}^g$ defined in  equation(140) is obtained. Accordingly, it is easy to verify that the right hand side expression of  equation(142) can be derived starting with the right hand side expression of  equation(85), with $\alpha, \beta~\gamma$ parameters replaced by $\lambda$ parameters, and then making the interchange $(1\leftrightarrows2)$.

Inserting  equation(25) in  equation(142), the closed-form expression for the generating integral $K_{45}^g$ is derived, from which all the limiting expressions for $K_{45}^g$ can be obtained, following the procedure adopted for deriving  equations(90-99) corresponding to $K_{41}^g$.

\textbf{Expression for $<r_{12}r_{13}/r_{34}>$}

An expression for $<r_{12}r_{13}/r_{34}>$ can be obtained making use of the generating integral $K_{45}^g$. A closed-form expression for the following integral
\begin{equation} %Equation No.143
\begin{split}
&K_{45}(-1, -1, -1, -1, ; 1, 1, -1; \lambda_1, \lambda_2, \lambda_3, \lambda_4 ; \lambda_{12}, \lambda_{13}, \lambda_{34}) \\
&= \int{d\overrightarrow{r_1}d\overrightarrow{r_2}d\overrightarrow{r_3}d\overrightarrow{r_4}} (r_1r_2r_3r_4)^{-1}(r_{12}r_{13}/r_{34}) \\
&\kern 3pc \times exp(-\lambda_1r_1 -\lambda_2r_2 -\lambda_3r_3 -\lambda_4r_4 -\lambda_{12}r_{12} -\lambda_{13}r_{13}- \lambda_{34}r_{34}),
\end{split}
\end{equation}
can be obtained  following similar procedure adopted for evaluating $ <r_{12}r_{23}/r_{34}>$, employing  equation(107). Substituting $\lambda_{12}=\lambda_{13}=\lambda_{34}=0$ in the closed-form expression for the above integral, the following integral is evaluated :
\begin{equation} %Equation No.144
\begin{split}
&K_{45}(-1, -1, -1, -1 ; 1, 1, -1; \lambda_1, \lambda_2, \lambda_3, \lambda_4 ; 0, 0, 0) = \frac{512}{3}\frac{\pi^4}{\lambda_2^2\lambda_4^2}\sum_{i=1}^8X_i(1\leftrightarrows2).
\end{split}
\end{equation}
Here,  $X_i(1\leftrightarrows2)$, $i=1,2,\cdots,8$, are the functions given by  equation(114) followed by the interchange $(1\leftrightarrows2)$. This integral has actually been evaluated and found to be consistent with the relation between $K_{41}^g$ and $K_{45}^g$.

\textbf{Expression for $<r_{12}r_{34}/r_{13}>$}

Expressions for the integral
\begin{equation} %Equation No.145
\begin{split}
&K_{45}(-1, -1, -1, -1, ; 1, -1, 1; \lambda_1, \lambda_2, \lambda_3, \lambda_4 ; \lambda_{12}, \lambda_{13}, \lambda_{34}) \\
&= \int{d\overrightarrow{r_1}d\overrightarrow{r_2}d\overrightarrow{r_3}d\overrightarrow{r_4}} (r_1r_2r_3r_4)^{-1}(r_{12}~r_{34}/r_{13}) \\
&\kern 3pc \times exp(-\lambda_1r_1 -\lambda_2r_2 -\lambda_3r_3 -\lambda_4r_4 -\lambda_{12}r_{12} -\lambda_{13}r_{13}- \lambda_{34}r_{34}),
\end{split}
\end{equation}
and, the integral obtained as a special case from  equation(145) with $\lambda_{12}=\lambda_{13}=\lambda_{34}=0$, can be derived as was done relating to  equations(90) and (103-106). In particular, a closed-form expression for $K_{45}^g(\lambda_1, \lambda_2, \lambda_3, \lambda_4; \lambda_{12}, 0, \lambda_{34})$ has been derived and shown to be exactly identical with the one obtained from the right hand side expression in  equation(90) by making interchange $(1\leftrightarrows2)$, as expected. Similarly, the expression derived for $K_{45}(-1, -1, -1, -1; 1, -1, 1; \lambda_1, \lambda_2, \lambda_3, \lambda_4; 0, 0, 0)$ has been shown to be exactly identical with the one obtained from the right hand side expression in  equation(105) by performing the interchange $(1\leftrightarrows2)$, as expected.

\textbf{(b) Analytic evaluation of $K_{46}^g$}

The generating integral $K_{46}^g$ is given by the equation
\begin{equation} %Equation No.146
\begin{split}
&K_{46}^g(\lambda_1, \lambda_2, \lambda_3, \lambda_4 ; \lambda_{12}, \lambda_{24}, \lambda_{43}) = \int{d\overrightarrow{r_1}d\overrightarrow{r_2}d\overrightarrow{r_3}d\overrightarrow{r_4}} (r_1r_2r_3r_4r_{12}r_{24}r_{43})^{-1}\\
&\times exp(-\lambda_1r_1 -\lambda_2r_2 -\lambda_3r_3 -\lambda_4r_4 -\lambda_{12}r_{12} -\lambda_{24}r_{24}- \lambda_{43}r_{43}),
\end{split}
\end{equation}
which can be rewritten as 
\begin{equation} %Equation No.147
\begin{split}
&K_{46}^g = \int{d\overrightarrow{r_2}d\overrightarrow{r_4}} (r_2r_4r_{24})^{-1} exp(-\lambda_2r_2 -\lambda_4r_4 -\lambda_{24}r_{24})\\
&~\times~J(\lambda_1, \lambda_{12},r_2)~ J(\lambda_3, \lambda_{43},r_4),
\end{split}
\end{equation}
where the J's are given by  equations(1) and (3). Replacing the J's by their closed-form expressions,  equation(147) simplifies to give
\begin{equation} %Equation No.148
\begin{split}
&K_{46}^g(\lambda_1, \lambda_2, \lambda_3, \lambda_4; \lambda_{12}, \lambda_{24},\lambda_{43})=\frac{16\pi^2}{(\lambda_1^2-\lambda_{12}^2)(\lambda_3^2-\lambda_{43}^2)}\\
&\times\Big[I_2(-2,-2,-1;\lambda_2+\lambda_{12}, \lambda_4+\lambda_{43}, \lambda_{24})-I_2(-2,-2,-1;\lambda_2+\lambda_{12}, \lambda_3+\lambda_4, \lambda_{24})\\
&-I_2(-2,-2,-1;\lambda_1+\lambda_2, \lambda_4+\lambda_{43}, \lambda_{24})+I_2(-2,-2,-1;\lambda_1+\lambda_2, \lambda_3+\lambda_4, \lambda_{24})\Big],
\end{split}
\end{equation}
where expressions for the $I_2$ integrals are given by  equation(25)

Comparing  equation(85) for the integral $K_{41}^g$ with  equation(148) for the integral $K_{46}^g$, it is observed that if a change $(3\leftrightarrows4)$ is made in the $\lambda$ subscripts in  equation(85), it leads to  equation(148). This is as expected.

Similarly, comparing  equation(148) for $K_{46}^g$ and  equation(136) for $K_{43}^g$, it is observed that the closed-form expression for the former can be obtained from that for the latter by the interchange $(1\leftrightarrows2)$ performed in the latter expression.

\textbf{(c) Analytic evaluation of $K_{47}^g$}

The generating integral $K_{47}^g$ is defined as 
\begin{equation} %Equation No.149
\begin{split}
&K_{47}^g(\lambda_1, \lambda_2, \lambda_3, \lambda_4 ; \lambda_{23}, \lambda_{31}, \lambda_{14}) = \int{d\overrightarrow{r_1}d\overrightarrow{r_2}d\overrightarrow{r_3}d\overrightarrow{r_4}} (r_1r_2r_3r_4r_{23}r_{31}r_{14})^{-1}\\
&\times exp(-\lambda_1r_1 -\lambda_2r_2 -\lambda_3r_3 -\lambda_4r_4 -\lambda_{23}r_{23} -\lambda_{31}r_{31}- \lambda_{14}r_{14}),
\end{split}
\end{equation}
which can be expressed as 
\begin{equation} %Equation No.150
\begin{split}
&K_{47}^g = \int{d\overrightarrow{r_1}d\overrightarrow{r_3}} (r_1r_3r_{31})^{-1} exp(-\lambda_1r_1 -\lambda_3r_3 -\lambda_{31}r_{31})\\
&~\times~J(\lambda_2, \lambda_{23},r_3)~ J(\lambda_4, \lambda_{14},r_1),
\end{split}
\end{equation}
with expressions for the J's given in  equation(3). Substituting these expressions in  equation(150), it simplifies to 
\begin{equation} %Equation No.151
\begin{split}
&K_{47}^g(\lambda_1, \lambda_2, \lambda_3, \lambda_4; \lambda_{23}, \lambda_{31},\lambda_{14})=\frac{16\pi^2}{(\lambda_2^2-\lambda_{23}^2)(\lambda_4^2-\lambda_{14}^2)}\\
&\times\Big[I_2(-2,-2,-1;\lambda_1+\lambda_{14}, \lambda_3+\lambda_{23}, \lambda_{31})-I_2(-2,-2,-1;\lambda_1+\lambda_{14}, \lambda_2+\lambda_3, \lambda_{31})\\
&-I_2(-2,-2,-1;\lambda_1+\lambda_4, \lambda_3+\lambda_{23}, \lambda_{31})+I_2(-2,-2,-1;\lambda_1+\lambda_4, \lambda_2+\lambda_3, \lambda_{31})\Big],
\end{split}
\end{equation}
with the $I_2$ integrals given by  equation(25).

If the right hand side expression in  equation(123) for $K_{42}^g$ is compared with the right hand side expression in  equation(151) for $K_{47}^g$, it is clearly observed that the latter can be obtained from the former if an interchange $(4\leftrightarrows1)$ is performed. This is consistent with the observation relating to the integrands for $K_{42}^g$ and $K_{47}^g$ defined in  equations(120) and (149), respectively.

Similarly it is verified that the right hand side expression in  equation(151) for $K_{47}^g$ can be obtained from the right hand side expression in  equation(139) for $K_{44}^g$ by making interchange $(2\leftrightarrows3)$ in  equation(139), as expected.

\textbf{(d)Analytic evaluation of $K_{48}^g$}

The generating integral $K_{48}^g$ is defined by the relation 
\begin{equation} %Equation No.152
\begin{split}
&K_{48}^g(\lambda_1, \lambda_2, \lambda_3, \lambda_4 ; \lambda_{23}, \lambda_{24}, \lambda_{41}) = \int{d\overrightarrow{r_1}d\overrightarrow{r_2}d\overrightarrow{r_3}d\overrightarrow{r_4}} (r_1r_2r_3r_4r_{23}r_{24}r_{41})^{-1}\\
&\times exp(-\lambda_1r_1 -\lambda_2r_2 -\lambda_3r_3 -\lambda_4r_4 -\lambda_{23}r_{23} -\lambda_{24}r_{24}- \lambda_{41}r_{41}),
\end{split}
\end{equation}
which can be rewritten as 
\begin{equation} %Equation No.153
\begin{split}
&K_{48}^g = \int{d\overrightarrow{r_2}d\overrightarrow{r_4}} (r_2r_4r_{24})^{-1} exp(-\lambda_2r_2 -\lambda_4r_4 -\lambda_{24}r_{24})\\
&~\times~J(\lambda_3, \lambda_{23},r_2)~ J(\lambda_1, \lambda_{41},r_4)\cdot
\end{split}
\end{equation}
Inserting  equation(3) in  equation(153), the following expression for $K_{48}^g$ is obtained :
\begin{equation} %Equation No.154
\begin{split}
&K_{48}^g(\lambda_1, \lambda_2, \lambda_3, \lambda_4; \lambda_{23}, \lambda_{24},\lambda_{41})=\frac{16\pi^2}{(\lambda_3^2-\lambda_{23}^2)(\lambda_1^2-\lambda_{41}^2)}\\
&\times\Big[I_2(-2,-2,-1;\lambda_2+\lambda_{23}, \lambda_4+\lambda_{41}, \lambda_{24})-I_2(-2,-2,-1;\lambda_2+\lambda_{23}, \lambda_4+\lambda_1, \lambda_{24})\\
&-I_2(-2,-2,-1;\lambda_2+\lambda_3, \lambda_4+\lambda_{41}, \lambda_{24})+I_2(-2,-2,-1;\lambda_2+\lambda_3, \lambda_4+\lambda_1, \lambda_{24})\Big],
\end{split}
\end{equation}
with expressions for the $I_2$ integrals given by  equation(25).

Comparing the right hand side expression in  equation(123) for $K_{42}^g$ with that in  equation(154) for $K_{48}^g$, it is observed that if a change $(3\leftrightarrows2)$ is performed in  equation(123), expression in  equation(154) is exactly reproduced, which is evident from integrands in  equations(120) and (152).

All the limiting expressions for the generating integrals $K_{46}^g$, $K_{47}^g$ and $K_{48}^g$
can be derived in the same manner as was done for establishing  equations(90-99) in the case of the generating integral $K_{41}^g$.\\
\hfill\break
\textbf{6. Conclusion}\\
\hfill\break
The intergals evaluated in this paper are likely to be utilized by those workers who do calculations employing Hy-CI and/or E-Hy-CI methods of variation. The programs sometimes developed by them for calculation of correlated integrals numerically can be tested with the exct values obtained from closed-form expressions of such integrals reported here.\\
\hfill\break
\textbf{Acknowledgments} \\
\hfill\break
The present contribution is dedicated to the memory of late Prof.D.K.Rai, B.H.U., Varanasi under whose supervision I did my Ph.D. and who breathed  his last on July 12, 2012 at the age of 69. I am extremely thankful to Prof.S.N.Thakur for his elder brotherly attitude toward me and encouragement to continue research studies after my superannuation. I am highly indebted to Prof.F.W.King for sending me the analytical part in his recent paper (reference [59]) prior to publication. I am extremely grateful to Prof.J.S.Sims and Dr. Maria Belen Ruiz for periodic correspondences and helpful suggestions. I greatly acknowledge the helpful discussions with Prof.S.K.Patra during the course of this work, and for going through the manuscript critically. I also greatly acknowledge the kind help rendered by the Director, IOP, Bhubaneswar by providing me library and computer facilities of the Institute. It is a pleasure to thank Rajesh for getting the manuscript typed nicely.

\textbf{References} 

\begin{small}
\begin{enumerate}[label={[\arabic*]}]
  \item Lowdin P -O 1959 \textit{Advances in Chemical Physics} ed I Prigogine (New York : Interscience) \textbf{2} pp 207 - 322
  \item Hylleraas E A 1929 Z. Physik \textbf{54} 347 
  \item Morse P M and Feshback H 1953 \textit{Methods of Theoretical Physics} (Inc. New York : Mc. Graw Hill Book Company) p 1738
  \item Li C, Wang L and  Yan Z -C 2013 Phys. Rev. A \textbf{88} 052513
  \item Drake G W F 1996 \textit{Atomic, Molecular and Optical Physics Handbook} ed G W F Drake (Wodbury NY : AIP) p 154
  \item Harris F E and Smith V H Jr 2005 Adv. Quantum Chem. \textbf{48} 407
  \item King F W, Quicker D and Langer J 2011 J. Chem. Phys. \textbf{134} 124114
  \item James H M and Coolidge AS 1936 Phys. Rev. \textbf{49} 688
  \item Roothan C C J 1951 J. Chem. Phys. \textbf{19} 1445
  \item King F W 1999 Adv. At. Mol. Opt. Phys. \textbf{40} 57
  \item Pelzl P J and King F W 1998 Phys. Rev. E \textbf{57} 7268
  \item Yan Z  -C and Drake G W F 1997 J. Phys. B : At. Mol. Opt. Phys. \textbf{30} 4723
  \item Frolov A M and Smith V H Jr 1997 Int. J. Quantum Chem. \textbf{63} 269
  \item Fromm D M and Hill R N 1987 Phys. Rev. A \textbf{36} 1013
  \item Remiddi E 1991 Phys. Rev. A \textbf{44} 5492
  \item Harris F E 1997 Phys. Rev. A \textbf{55} 1820
  \item Sims J S and Hagstrom S A 2003 Phys. Rev. A \textbf{68} 016501
  \item Sims J S and Hagstrom S A 2003 Phys. Rev. A \textbf{68} 059903 (E)
  \item Harris F E, Frolov A M and Smith V H Jr 2004 Phys. Rev. A \textbf{69} 056501
  \item Pachucki K, Puchalski M and Remiddi E 2004 Phys. Rev. A \textbf{70} 032502
  \item Puchalski M and Pachucki K 2006 Phys. Rev. A \textbf{73} 022503
  \item Puchalski M and Pachucki K 2010 Phys. Rev. A \textbf{81} 052505
  \item Puchalski M, Kedziera D and Pachucki K 2011 Phys. Rev. A \textbf{84} 052518
  \item Harris F E 2005 Int. J. Quantum Chem. \textbf{105} 857
  \item Harris F E 2009 Phys. Rev. A \textbf{79} 032517
  \item Sims J S and Hagstrom S A 1971 Phys. Rev. A \textbf{4} 908
  \item Sims J S and Hagstrom S A 1971 J. Chem. Phys. \textbf{55} 4699
  \item Burke E A 1963 Phys. Rev. \textbf{130} 1871
  \item Gentner R F and Burke E A 1968 Phys. Rev. \textbf{176} 63
  \item Perkins J F 1969 J. Chem. Phys. \textbf{50} 2819
  \item Perkins J F 1973 Phys. Rev. A \textbf{8} 700
  \item Sims J S and Hagstrom S A 1975 Phys. Rev. A \textbf{11} 418
  \item Pipin J and Bishop D M 1992 Phys. Rev. A \textbf{45} 2736
  \item Sims J S and Hagstrom S A 2009 Phys. Rev. A \textbf{80} 052507
  \item Sims J S, Hagstrom S A, Munch D and Bunge C F 1976 Phys. Rev. A \textbf{13} 560
  \item Clay D C and Handy N C 1976 Phys. Rev. A \textbf{14} 1607
  \item Sims J S and Hagstrom S A 2002 Int. J. Quantum Chem. \textbf{90} 1600
  \item Ruiz M B, Margraf J T and Frolov A M 2013 Phys. Rev. A \textbf{88} 012505
  \item Sims J S and Hagstrom S A 2011 Phys. Rev. A \textbf{83} 032518
  \item Sims J S and Hagstrom S A 2014 J. Chem. Phys. \textbf{140} 224312
  \item Sims J S and Hagstrom S A 2004 J.Phys. B: At Mol. Opt. Phys. \textbf{37} 1519
  \item Sims J S and Hagstrom S A 2007 J. Phys. B: At. Mol. Opt. Phys. \textbf{40} 1575
   \item Sims J S and Hagstrom S A 2015 J. Phys. B: At. Mol. Opt. Phys. \textbf{48} 175003
   \item Ruiz M B 2009 J. Math. Chem. \textbf{46} 24            
	\item Ruiz M B 2009 J. Math. Chem. \textbf{46} 1322
		\item Ruiz M B 2011 J. Math. Chem. \textbf{49} 2457
		\item Ruiz M B 2016 J. Math. Chem. \textbf{54} 1083
		\item Frolov A M, Ruiz M B and Wardlaw D M 2014 Chem. Phys. Lett. \textbf{608} 191
		\item King F W 1993 J. Chem Phys. \textbf{99} 3622
		\item Harris F E, Frolov A M and Smith V H Jr 2004 J. Chem. Phys. \textbf{120} 3040
		\item King F W 2004 J. Chem. Phys. \textbf{120} 3042 
		\item King F W 2014 J. Phys. B: At. Mol. Opt. Phys. \textbf{47} 025003
		\item Frolov A M 2008 J. Phys. B: At. Mol. Opt. Phys. \textbf{41} 059801
		\item Frolov A M 2004 J. Phys. B : At. Mol. Opt. Phys. \textbf{37} 2103
	\item Li C, Wang L and Yan Z  -C 2013 Int. J. Quantum Chem. \textbf{113} 1307
	\item Wang C, Mei P, Kurokawa Y, Nakashima H and Nakatsuji H 2012 Phys. Rev. A  \textbf{85} 042512
	\item Padhy B 2012 Asian J. Spectrosc. \textbf{Special Issue} p 157 ; arXiv: 1609.00269
	\item Padhy B 2013 Orissa J. Phys. \textbf{20(1)} p 11  ; arXiv: 1609.00112
	\item King F W 2016 J. Phys. B: At. Mol. Opt. Phys. \textbf{49} 105001
	\item Roothan C C J and Weiss A W 1960 Rev. Mod. Phys. \textbf{32} 194
	\item Calais J  -L and Lowdin P  -O 1962 J. Mol. Spectrosc. \textbf{8} 203
	\item Harris F E, Frolov A M and Smith V H Jr 2004 J. Chem. Phys. \textbf{121} 6323
	\item Bonham R A 1965 J. Mol. Spectrosc. \textbf{15} 112
	\item Bonham R A 1966 J. Mol. Spectrosc. \textbf{20} 197

	\item Roberts P J 1965 J. Chem. Phys. \textbf{43} 3547
	\item Roberts P J 1967 J. Chem. Phys. \textbf{47} 3662
	\item Roberts P J 1968 J. Chem. Phys. \textbf{49} 2954
	\item Harris F E, Frolov A M and Smith V H Jr 2004 J. Chem. Phys. \textbf{120} 9974
	\item Coolidge A S and James H M 1937 Phys. Rev. \textbf{51} 855
	\item Abramowitz M and Stegun I A (ed) 1972 \textit{Handbook of Mathematical Functions} (New York : Dover) p 230
	\item Sack R A 1964 J. Math. Phys. \textbf{5} 245
	\item Perkins J F 1968 J. Chem. Phys. \textbf{48} 1985
  
\end{enumerate}
\end{small}

\end{document}